\documentclass[reprint,3p,times,onecolumn,12pt]{elsarticle}
\pdfoutput=1 
\usepackage{geometry}
\usepackage{mathtools}
\usepackage{graphicx}
\usepackage{latexsym}
\usepackage{amssymb}
\usepackage{amscd,amssymb}
\usepackage[arrow,matrix]{xy}
\usepackage{graphicx}
\usepackage{epstopdf}

\biboptions{sort&compress}

\usepackage{ulem}
\usepackage{setspace}

\usepackage{amscd,amsmath,amsthm,amssymb}
\usepackage{amsmath,amssymb}
\usepackage[colorlinks=true,bookmarks=false,citecolor=blue,urlcolor=blue]{hyperref} %pdflatex

\newcommand{\bea}{\begin{eqnarray}}
\newcommand{\eea}{\end{eqnarray}}
\newcommand{\bes}{\begin{subequations}}
\newcommand{\ees}{\end{subequations}}

\begin{document}
	\title{Manipulation of vector solitons in a system of inhomogeneous coherently coupled nonlinear Schr{\"o}dinger models\\ with variable nonlinearities}
	
	\author[rb]{\large R. Babu Mareeswaran}
	\author[ks1,ks2]{\large K. Sakkaravarthi \corref{cor}}
	\author[tk]{\large T. Kanna\corref{cor}}
	
	\address[rb]{\it Department of Physics, PSG College of Arts and Science, Coimbatore--641 014, India}
	\address[ks1]{Department of Physics, National Institute of Technology, Tiruchirappalli -- 620 015, Tamil Nadu, India}
	\address[ks2]{Centre for Nonlinear Dynamics, School of Physics, Bharathidasan University, Tiruchirappalli -- 620 024, India}
	\address[tk]{\it Nonlinear Waves Research Laboratory, PG and Research Department of Physics,\\ Bishop Heber College (Affiliated to Bharathidasan University), Tiruchirapalli--620 017, Tamil Nadu, India\vspace{-1.0cm}}
	
	\cortext[cor]{Corresponding authors. \newline Email address: babu\_nld@rediffmail.com; ksakkaravarthi@gmail.com; kanna\_phy@bhc.edu.in}
	
	\journal{\bf Journal of Physics A: Mathematical and Theoretical}
	\date{}
 
\begin{abstract}
	We investigate non-autonomous solitons in a general coherently coupled nonlinear Schr\"odinger (CCNLS) system with temporally modulated nonlinearities and with an external harmonic oscillator potential. This general CCNLS system encompasses three distinct types of CCNLS equations that describe the dynamics of beam propagation in an inhomogeneous Kerr-like nonlinear optical medium for different choices of nonlinear polarizations owing to the anisotropy of the medium. We identify a generalized similarity transformation to relate the considered model into the standard integrable homogeneous coupled nonlinear evolution equations with constant nonlinearities, accompanied by a constraint relation expressed in the form of the Riccati equation. With the help of a non-standard Hirota's bilinearization method and exact soliton solutions, we explore the impact of varying nonlinearities and refractive index in the propagation and collisions analytically by reverse engineering. Interestingly, we show the emergence of several modulated solitonic phenomena such as  periodic oscillation, amplification, compression, tunneling/cross-over, excitons, as well as their combined effect in the single-soliton propagation and two-soliton collisions with appropriate forms of nonlinearity. Notably, we identify a tool to transform the nature of soliton collisions with certain type of inhomogeneous nonlinearities. The results could be of significant interest to the studies on management of nonlinear waves in the contexts like nonlinear optics and can also be extended to Bose-Einstein condensates and super-fluids.
	
	\begin{keyword}{Nonlinear optics; Inhomogeneous system; Coupled nonlinear Schr\"odinger equations; Vector solitons; Variable nonlinearity; Soliton management.\newline}
		------------------------------------------------------------------------------------------------------------------------\newline
\noindent {\large Reference: {\it Journal of Physics A: Mathematical and Theoretical} {\bf 53} (2020).\newline
	 \url{https://doi.org/10.1088/1751-8121/abae3f}}
	\end{keyword}
\end{abstract}

\maketitle

\setstretch{1.25}
\section{Introduction}
Nonlinear wave phenomena have a deep physical and mathematical interest as they arise in a wide landscape ranging from science to engineering and technology such as nonlinear optics, fluid dynamics, plasma physics, lattice dynamics, and Bose-Einstein condensates (BECs) \cite{kiv-book,ablowitz,wiley,pana}. In nonlinear dynamical systems, introducing inhomogeneous and non-autonomous nonlinearities showcase distinct dynamical behaviour of nonlinear waves that find applications in optical communication, water waves, and Bose-Einstein Condensates (BECs) \cite{expoptics,boaris,optexp1,expoptics1,becexp2,becexp2a,becexp2b}. Along this direction, theoretical and experimental investigations have been carried out during the past few decades. {Recently, the spatially modulated Kerr nonlinearity is observed experimentally in nonlinear optics} \cite{optexp1,expoptics1}. In the context of BECs, the spatial (or) time modulated nonlinearity is achieved through the Feshbach resonance mechanism with a non-uniform magnetic field \cite{bloch}. Various reports on these types of inhomogeneous and non-autonomous systems are available in the literature. 

Among several types of nonlinear waves, solitons found ever increasing interest due to their remarkable stability and intriguing collision dynamics. Further, they have multifaceted applications in almost all areas of science and technology \cite{Peyrad}. Solitons emerging from nonlinear Schr\"odinger (NLS) type equation with temporally varying/distributed coefficients (dispersion/nonlinearity), so-called “non-autonomous soliton”, found important advancements in the context of optical communication systems. Nowadays, these non-autonomous solitons play an important role in optical fibre system (see \cite{1a,1a2,1a3,1a4,1a5,1b} and reference therein). In the pioneering works \cite{serkin1,serkin2}, the concept of non-autonomous soliton was first introduced within the framework of the NLS system with variable dispersion and nonlinear coefficients and it has been show that the amplitude, widths, velocity and central position of soliton was completely affected by varying management parameters (dispersion and nonlinearity). The bright and dark spatial self-similar solitons in graded-index fiber with linear refractive index have been investigated \cite{1c}. In ref. \cite{maha,Abdul}, authors have studied the dynamics of bright soliton in a dispersion managed erbium doped inhomogeneous fibre with gain/loss. Especially, the existence of bright-dark dispersion managed soliton with randomly varying birefringence has been investigated \cite{2a}. The dynamics of spatio-temporal light bullets in three-dimensional nonlocal NLS system with variable coefficients have also been studied \cite{3} and it shows that intensity, width, phase and the chirp of light bullets are strongly modified by dispersion and nonlinearity coefficients. Specifically, the spatio-temporal multi-soliton solutions with and without continuous wave backgrounds in the NLS equation with variable dispersion and nonlinearity coefficients have been obtained in \cite{4}. Moreover, the snake-like nonautonomous solitons in planar grating waveguides have been investigated \cite{5a,5b} and reported the control of the shape preserved soliton's motion on the graded-index waveguides. Recently, the shape of the dissipative dispersion-managed solitons in optical fiber systems with lumped amplification have been studied both experimentally and numerically \cite{6a}. 
Apart from the solitons, analysis on the exact periodic traveling wave and soliton pair (bright-dark and dark-dark) solutions of the coupled NLS equations with harmonic potential and variable coefficients has been reported \cite{6b}. The effects of the periodically modulated nonlinearity on the soliton propagation and interaction in a dispersion-managed birefringence system is also of much importance \cite{6c,similariton,prebabu}. Analytical vector non-autonomous soliton solutions for the coupled NLS with spatially modulated coefficients and coherent coupling were studied \cite{7}. Particularly, the phase dynamics of bright and dark solitons in variable coefficient coupled NLS equation was reported \cite{7a}. More recently, the evolution and stability of bright vector soliton in coupled Ginzburg-Landau equation with variable coefficients has been studied \cite{8}. This clearly indicates that study of non-autonomous solitons is of much importance not only in one dimensional NLS type systems but also in higher dimension and dissipative nonlinear systems as well as multi-component optical systems such as multi-mode fibers. 

Based on the above interesting works, in this work, we give a detailed analysis of soliton management in a general coupled NLS type system with variable Kerr nonlinearity and linear parabolic refractive index profile arising in the context of polarized light beam propagation in low birefringent nonlinear optical media. Here the relative phase factors of the co-propagating electric fields lead to the onset of four–wave mixing effects. This coupling arises naturally in weakly anisotropic or birefringent media. The following system of coherently coupled nonlinear Schr\"odinger (CCNLS) equations (in dimensionless form) \cite{kiv-book,akhm-book} describes such type of beam propagation in Kerr-like nonlinear medium:
\bes\bea
\hspace{-1.4cm}i\frac{\partial A_1}{\partial z}+\left(\delta\frac{\partial^2 }{\partial x^2} +\upsilon(x,z)\right)A_1+\gamma(z)[\sigma_{11}|A_{1}|^2+\sigma_{12}|A_{2}|^2]A_{1}-\delta_1\gamma(z) A_2^2A_{1}^*=0,\\
\hspace{-1.4cm}i\frac{\partial A_2}{\partial z}+\left(\delta\frac{\partial^2 }{\partial x^2} +\upsilon(x,z)\right)A_2+\gamma(z)[\sigma_{21}|A_{1}|^2+\sigma_{22}|A_{2}|^2]A_{2}-\delta_2\gamma(z)A_1^2A_{2}^*=0, 
\eea\label{nccnls}\ees
where $A_{j}$~($j=1,2$) are the slowly varying envelopes of the electric fields associated with two orthogonally polarized components. In Eq.~(1),  $x$ and $z$ are the transverse and longitudinal coordinates, respectively, while the asterisk ($*$) denotes the complex conjugation and $\delta$ represents the group velocity dispersion coefficient \cite{kerrmedia}. Further, $\gamma(z)$ is the variable nonlinear parameter, the $z$-dependence of which stems from the inhomogeneity of the optical medium \cite{boaris,expoptics1} and $\upsilon(x,z)$ is the graded refractive index profile in the form \cite{kivshar}. Eq.~(1) contains phase-independent incoherent nonlinearities given by  $\sigma_{ij},~i,j=1,2$, where $\sigma_{11}$ and $\sigma_{22}$ represent the strengths of self-phase modulation (SPM), while $\sigma_{12}$ and $\sigma_{21}$ denote the strengths of cross-phase modulation (XPM). Additionally, the four-wave mixing (FWM) nonlinearity strength is represented by the coefficients $\delta_1$ and $\delta_2$ of the phase-dependent coherent coupling. Here one can have a different consideration with varying dispersion effect or dispersion management $\delta(z)$ as well as spatio-temporal modulated nonlinearities $\gamma(x,z$) in the CCNLS system (1), which deserve a separate intensive investigation and shall be reported shortly. 
For a special set of parameters ($\sigma_{12}=\sigma_{21}=\delta_j=0$) along with constant nonlinearities and in the absence of graded refractive index, the system (1) reduces to the standard scalar NLS equation for the two separate modes $A_1$ and $A_2$, namely $iA_{j,t}+\delta A_{j,xx}+\gamma |A_j|^2A_j=0,~j=1,2$, whose various nonlinear wave solutions including solitons, breathers, rogue waves, periodic waves under different physical situations have been extensively studied. Further, when the FWM nonlinearity is absent, the soliton dynamics for focusing and mixed type nonlinear Schr\"odinger modles were investigated with explicit soliton solutions and asymptotic analysis exploring the energy-sharing and elastic collisions of bright solitons \cite{RK97,akhm,tkopt,tkopt2,tkopt3}. 

Moreover, it is important to mention that Eq. (1) is a general version of inhomogeneous CCNLS system. Interestingly, it results into six different models for appropriate choices of dispersion and nonlinearity coefficients. The inhomogeneous model (1) with constant nonlinearities and refractive index describes beam propagation in uniform/homogeneous Kerr-like nonlinear medium reads as 
\bes\bea
i\frac{\partial A_1}{\partial z}+\delta\frac{\partial^2 A_1}{\partial x^2}+\gamma\left(\sigma_{11} |A_{1}|^2+\sigma_{12} |A_{2}|^2\right)A_{1}-\delta_1\gamma A_2^2A_{1}^*=0,\\
i\frac{\partial A_2}{\partial z}+\delta\frac{\partial^2 A_2}{\partial x^2}+\gamma\left(\sigma_{21} |A_{1}|^2+\sigma_{22} |A_{2}|^2\right)A_{2}-\delta_2\gamma A_1^2A_{2}^*=0.
\eea\label{gen-2ccnls}\ees
The above Eq.~(2) is non-integrable in general and shall pass the integrability test only for certain choices of coefficients \cite{Park}. As mentioned earlier, the above generalized two-component CCNLS equation shall take different versions for various choices of nonlinearity and dispersion coefficients, as given below.
\bea
&\mbox{Model (i) ~~\cite{tkjpa}:}  & 
\delta=-1, \sigma_{11}=\sigma_{22}=-1, \sigma_{12}=\sigma_{21}=-2, \delta_1=\delta_2=1, \label{ccnls10}\\
&\mbox{Model (ii) ~\cite{ksjpa}:}  &
\delta=1, ~\sigma_{11}=\sigma_{22}=1, \sigma_{12}=\sigma_{21}=2, \delta_1=\delta_2=1, \label{ccnls11}\\
&\mbox{Model (iii) \cite{tk2011ncnsd}:}  &
\delta=-1, \sigma_{11}=\sigma_{22}=-1, \sigma_{12}=\sigma_{21}=-2, \delta_1=\delta_2=-1,  \label{ccnls11a}\\
&\mbox{Model (iv) \cite{ksjmp}:}  &
\delta=1, \sigma_{11}=-\sigma_{22}=1, \sigma_{12}=-\sigma_{21}=-2, \delta_1=-\delta_2=1, \label{ccnls13}\\
&\mbox{Model (v) \cite{tkpla16}:}  &
\delta=1, ~ \sigma_{11}=\sigma_{22}=-1, \sigma_{12}=\sigma_{21}=-2, \delta_1=\delta_2=1, \label{ccnls16a}\\
&\mbox{Model (vi) \cite{tkpla16}:}  &
\delta=1, ~ \sigma_{11}=\sigma_{22}=1, \sigma_{12}=\sigma_{21}=2, \delta_1=\delta_2=-1. \label{ccnls16b}
\eea
Among the above six versions, models (v) and (vi) correspond to the choices (\ref{ccnls16a}) and (\ref{ccnls16b}) are equivalent to (\ref{ccnls10}) when $x\rightarrow ix$ and $z\rightarrow -z$, respectively, while the model (iii) arising for the choice (\ref{ccnls11a}) can be reduced to model (ii) resulting for (\ref{ccnls11}) when $z\rightarrow -z$. Further, based on the impact of nonlinearities, models resulting for the choices (\ref{ccnls10}), (\ref{ccnls16a}) and  (\ref{ccnls16b}) shall be defined as CCNLS systems with positive coherent coupling, while that of the choices (\ref{ccnls11}) and  (\ref{ccnls11a}) are as CCNLS systems with negative coherent coupling, and the choice (\ref{ccnls13}) is designated to the CCNLS system with mixed type nonlinearities. Thus, effectively there exist only three distinct models (\ref{ccnls10}), (\ref{ccnls11}), and (\ref{ccnls13}) representing the coherent propagation of two orthogonally polarized modes featuring different SPM, XPM, and four-wave mixing nonlinearities. The exact bright soliton solutions of these models (\ref{ccnls10}), (\ref{ccnls11}), and (\ref{ccnls13}) were constructed by employing a non-standard way of Hirota's bilinearization method in Refs. \cite{tkjpa}, \cite{ksjpa}, and \cite{ksjmp}, respectively, along with a detailed study on the propagation and collision dynamics of these solitons. Particularly, the solitons were classified as coherently coupled solitons and incoherently coupled solitons based on the presence and absence of the phase-dependent (four-wave mixing) nonlinearity, respectively, exhibiting single-hump, double-hump and flat-top profiles. Further, an interesting energy-switching collision of bright coherent-incoherent solitons was investigated in addition to their energy-sharing and elastic collisions \cite{tkjpa,ksjpa,ksjmp}.   

On the other hand, equations similar to (3-8) can also arise in the context of BECs governing the dynamics of spinor condensates when the spin-mixing nonlinearity plays a crucial role. For example, soliton solutions and their interactions of an autonomous and non-autonomous spin-1 condensate system usually referred as three-coupled Gross-Pitaevskii equations were obtained in Refs. \cite{Wadati-spin,Wadati-spin2,Wadati-spin3,Theis,tkwcna,tkpla14} and have been classified as interesting ferromagenetic and polar solitons, based on the effect of spin-mixing nonlinearity. This three-component system reduces to the two-component CCNLS type system when we consider pseudo-spinors and is referred as degenerate CCNLS system \cite{tkpla16}. A variant degenerate coherently coupled spin system similar to Eq. (7-8) has been discussed in autonomous and non-autonomous settings \cite{tkpla16}. The study unraveled various coherent structures through linear superposition and by varying nonlinearities and external potential. 

Motivated by the propagation and collision dynamics of solitons in homogeneous optical media \cite{tkjpa,ksjpa,ksjmp} as well as the investigation of non-autonomous solitons in BECs \cite{tkpla14}, in this work, we focus our investigation on the generalized CCNLS system (1)  with varying nonlinearity and refractive index profile. The objectives are to construct exact soliton solutions with inhomogeneity and varying nonlinearity coefficients, which can be implemented through an appropriate similarity or lens type transformation. Further, the propagation as well as collision dynamics of such inhomogeneous solitons will be studied extensively with appropriate analysis and graphical demonstrations.  

The remaining part of this work is arranged in the following manner: The conversion of inhomogeneous 2-CCNLS system (1) into a homogeneous version (2) with a similarity transformation is given in Sec. \ref{trans}. along with the importance of considered varying nonlinearities. Section \ref{sec-1sol} consists of the propagation dynamics of inhomogeneous solitons which enacts the manipulation mechanism of optical solitons through nonlinear optical fibers/communication systems. Further, various types of inhomogeneous soliton collisions are presented in Sec. \ref{sec-2sol} with categorical analysis. Further, the possibility and occurrence of inhomogeneous soliton bound states are discussed in Sec. \ref{sec-bound}. The final section  \ref{sec-conc} is devoted to summarize the important results along with certain future perspectives.

\section{Transformation to the Integrable Homogeneous CCNLS Model}\label{trans}
Solving inhomogeneous nonlinear models is comparatively difficult, but quite possible, than that of their homogeneous (constant parameter) counterparts. However, we require explicit solutions for a complete understanding of the considered system. This can be accomplished by two broad routes: directly solving the equations by retaining the variable coefficients and transforming the equation into a convenient model that can be exactly solved with various analytical methods. 

In this section, we adopt the second route by implementing a similarity transformation to extract explicit solutions for Eq.~(1). For this purpose, we apply the following similarity transformation to Eq.~(1):
\bea
&&A_j(x,z) = \rho(z)~Q_j(X(x,z),Z(z))~\text{exp}[i\zeta(x,z)],\quad j=1,2.
\label{simi}
\eea
where $\rho(z)$ is the amplitude, while $\zeta(x,z)$ is the phase and $X(x,z)$ and $Z(z)$ are the similarity variables, the explicit form of all these variables has to be determined. The above transformation (9) reduces Eq. (1) into the following homogeneous CCNLS equation:
\bes\bea
&iQ_{1,Z}+\delta Q_{1,XX}+(\sigma_{11}|Q_1|^2+\sigma_{12}|Q_2|^2)Q_1 - \delta_1 Q_2^2Q_1^*=0,\\
&iQ_{2,Z}+ \delta Q_{2,XX}+(\sigma_{21}|Q_1|^2+\sigma_{22}|Q_2|^2)Q_2-\delta_2 Q_1^2Q_2^*=0.
\eea\label{ccnls}
\ees
The only difference between Eqs. (2) and (10) is that the constant nonlinearity coefficient $\gamma$ takes a fixed value of $\gamma=1$ in the latter, while it stays arbitrary in the former. To determine the unknown functions of (9), we substitute it into (1) from which we obtain a set of partial differential equations (PDEs) for these unknown functions as given below.
\bes\bea
&&X_{xx}=0,\qquad ~\qquad \qquad
X_z+2\delta X_x {\zeta_x}=0,\qquad
\delta {\zeta_x^2}+\zeta_z-\frac{1}{2}F(z)x^2=0, \\
&&Z_z-{\rho^2(z)}\gamma(z)=0,\qquad
{ \rho_z}+\delta {\rho(z)} {\zeta_{xx}}=0,\qquad
X_x^2-{\rho^2(z)}\gamma(z)=0.
\eea\ees
Solving the above PDEs successively, we find the following expressions with auxiliary  real arbitrary constants $\epsilon_1$ and  $\epsilon_2$:
\bes\bea
&&\zeta(x,z) = -\frac{1}{4\delta}\frac{d}{dz}(\ln \gamma)x^2 +  \epsilon_1^2 \epsilon_2 {\gamma} x - \delta\epsilon_2^2 \epsilon_1^4\int {\gamma}^2 dz,\\
&&X(x,z) = ~\epsilon_1 \left({\gamma} x - 2\delta \epsilon_2 \epsilon_1^2\int {\gamma}^2 dz\right),\\
&&Z(z) = \epsilon_1^2 \int {\gamma}^2 dz,\\
&&\rho(z)=\epsilon_1 \sqrt{\gamma(z)},\\
&&F(z)=\frac{1}{\delta}\left(\frac{\gamma_z^2}{\gamma^2}-\frac{\gamma_{zz}}{2\gamma}\right).\label{potF}
\eea\label{str}\ees 
Note that the external potential and nonlinearity shall not be independent, either one can be arbitrary function of $z$, while the other admits suitable form through Eq. (12e). This condition can also be rewritten in a more convenient form of Riccati equation $Y_z-Y^2+\delta F(z)=0$, where $Y=\gamma_z/{\gamma}$. Now, with the identified explicit similarity transformation (9) and corresponding variables (12), one can easily construct exact nonlinear wave solutions of inhomogeneous model (1) when we are able to provide the respective solutions for constant parameter CCNLS equation (10). 

The above mentioned similarity transformation and the varying nonlinearities can be adopted for any inhomogeneous nonlinear systems, in the present case for all versions of the CCNLS models. Hereafter, we explore their impact in the propagation and collision dynamics of bright solitons of these CCNLS systems by constructing their explicit solutions obtained by using a non-standard type of Hirota's bilinearization method. We refrain from presenting the detailed procedure here and one can refer to \cite{tkjpa,ksjpa,ksjmp} for the systematic construction as well as the analysis on homogeneous solitons. Especially, we consider the three distinct versions (3), (4), and (6) one by one. 

\section{Inhomogeneous bright one-soliton}\label{sec-1sol}
In order to achieve the first objective, understanding the role of varying nonlinearities on soliton propagation, we construct explicit soliton solutions by adopting the Hirota's bilinearization method with an auxiliary function to homogeneous form (2) \cite{tkjpa,ksjpa,ksjmp} and deduce solutions of the inhomogeneous equations (1) under investigation, using (9). First of all, the bilinearizing transformation and generalized bilinear forms of CCNLS equation (2) can be written as,
\bes\bea
\hspace{-2.30cm}\mbox{Bilinear transformation} \Rightarrow &&Q_1=\frac{G}{F}, \qquad Q_2=\frac{H}{F},\\
\hspace{-2.30cm}\mbox{Bilinear equations \qquad} \Rightarrow &&(iD_Z+\delta D_X^2)G\cdot F= \delta SG^*,\\
&&(iD_Z+\delta D_X^2)H\cdot F= \delta_2 SH^*,\\
&&\delta D_X^2 F \cdot F = 2(|G|^2 +\delta_2 |H|^2),\\
&& S\cdot F = G^2 + \delta H^2,
\eea \label{bilinear}\ees 
where $G$, $H$ and $S$ are complex functions, while $F$ is a real function and $D$ represents the standard Hirota derivative \cite{Hirota-book} of the respective independent variables $Z$ and $X$. The above bilinear forms are applicable to the general CCNLS model (\ref{nccnls}) which includes all versions given by (3-8) for respective choices. When we apply the bilinearizing transformation (13a) to the three distinct versions of CCNLS models (3), (4) and (6), we shall get three different sets of bilinear forms. By combing those three forms, we have written in the above form (13) in a conenient way. By following the standard procedure, we can construct soliton solutions after expanding the dependent functions as power series and then by recursively solving the resultant ordinary differential equations arising at different orders of expansion parameters \cite{tkjpa,ksjpa,ksjmp}.

The generalized bright one-soliton solution of the inhomogeneous CCNLS system (1), especially, to the three distinct CCNLS Eqs. (3), (4), and (6), can be obtained as 
\bes\bea
&&A_j(x,z) = \frac{\epsilon_1 \sqrt{\gamma} \left(\alpha^{(j)}_1 e^{\eta_{1}}+e^{2\eta_{1}+\eta^{*}_1+\delta^{(j)}_{11}}\right)}{1++e^{\eta_{1}+\eta^{*}_1+R_1}+e^{\eta_{1}+\eta^{*}_1+\epsilon^{(j)}_{11}}} e^{i \zeta(x,z)}, \quad j=1,2,
\eea
\noindent where 
\bea
\hspace{-1.984cm}e^{\delta^{(1)}_{11}}= 
\frac{ \alpha^{(j)*}_1 S_1 }{2(k_1+k^*_1)^2}, e^{\delta^{(2)}_{11}}= 
\frac{\delta \delta_2  \alpha^{(2)*}_1 S_1}{2(k_1+k^*_1)^2}, e^{R_1} = \frac{ |\alpha^{(1)}_1|^2+\delta_2 |\alpha^{(2)}_1|^2}{(k_1+k^*_1)^2}, e^{\epsilon_{11}}=\frac{|S_1|^2}{(k_1+k^*_1)^4},~~~~\\
\eta_{1} = k_1(X+i\delta k_1Z), X= ~\epsilon_1 \left({\gamma} x - 2\delta \epsilon_2 \epsilon_1^2\int {\gamma}^2 dz\right),\quad Z = \epsilon_1^2 \int {\gamma}^2 dz.\qquad 
\eea 
The above soliton solution is obtained with an auxiliary function $S=S_1\mbox{e}^{2\eta_1}$, which plays an important role in the classification of the respective solitons as coherently- and incoherently-coupled solitons, and it takes the form 
\bea
&&S_1 = (\alpha^{(1)}_1)^2+\delta (\alpha^{(2)}_1)^2.
\eea \label{one-sol}\ees 
The above solution is applicable to all the three inhomogeneous CCNLS models (1) corresponding to (3), (4), and (6) with appropriate choices of dispersion ($\delta$), incoherent nonlinearities ($\sigma_{ij}$), and coherent nonlinearity ($\delta_j$). 

\subsection{Inhomogeneous Incoherently Coupled Solitons}
From the above general one-soliton solution (14), one can easily understand that the choice $S_1=0$ leads to $e^{\delta^{(1)}_{11}}$ and $e^{\epsilon_{11}}$ vanish. This results into a simple form corresponding to that of Manakov type solitons without the contribution from coherent nonlinearity, which can be designated as inhomogeneous incoherently coupled solitons (IICSs). Such type of inhomogeneous soliton can be casted in a standard hyperbolic form as 
%Based on the similarity transformations (\ref{str}), we obtain the soliton solution of system (\ref{nccnls}).
\bea
A_j(x,z)=B_j ~\text{sech}(\eta_{1R}+{R_1}/{2}) e^{i(\zeta+\eta_{1I})}, \quad j=1,2,
\label{1sol-iics}
\eea
\noindent where $B_j= \frac{1}{2}\alpha^{(j)}_1\epsilon_1 \sqrt{\gamma (z)} e^{-R_1/2}$, $\eta_{1R} = k_{1R}(X-2\delta k_{1I}Z)$, $\eta_{1I} = k_{1I}X+\delta (k_{1I}^2-k_{1R}^2)Z$,  
%\bea &&B_j= \frac{1}{2}\alpha^{(j)}_1\epsilon_1 \sqrt{\gamma (z)} e^{-R_1/2}, \quad
%~~~~~ \chi = \left(\eta_{1R}+\frac{R_1}{2}\right),~~~ \\
%  \eta_{1R} = k_{1R}(X-2k_{1I}Z),\quad  \eta_{1I} = k_{1I}X+(k_{1I}^2-k_{1R}^2)Z,
%~~~ \text{and} ~~~R_1 = \text{log}\left(\frac{k_{11}}{2k_{1R}}\right).
%\eea\ees
and other parameters $\zeta$, $X$ and $Z$ are as given in Eq.~(14). This IICS admits well-known bell-type/symmetric-single-hump profile with certain amplitude $B_j$, width (proportional to inverse of amplitude), and velocity $2\delta k_{1I}$ of propagation represented by the above form. Here we should note that the above IICS is possible for the models (3) and (4), while it results in singular solutions in system (6) as the restriction $S=0$ makes the denominator of general solution to vanish. 

\subsection{Inhomogeneous Coherently Coupled Solitons}
In addition to the above special/restricted IICS (15), the general bright one-soliton solution (14) usually contains the contribution from both coherent and incoherent nonlinearities ($S_1\neq 0$), which can be referred as inhomogeneous coherently coupled solitons (ICCSs). We can rewrite the solution (14) in a convenient form as below.
\bea
\hspace{-1.4cm}	A_j(x,z)=C_j\left(\frac{\mbox{cos}(P_j)~\mbox{cosh}\left(Q\right)
	+i~\mbox{sin}(P_j)~\mbox{sinh}(Q)}{4 \mbox{cosh}^2(Q)+L}\right)e^{i(\zeta+\eta_{1I})},~~ j=1,2, \label{1sol-iccs}
\eea
where $C_j=2\epsilon_1\sqrt{\gamma}~e^{{\frac{l_j+\delta_{11}^{(j)}-\epsilon_{11}}{2}}}$, $P_j=e^\frac{\delta_{11I}^{(j)}-l_{jI}}{2}$, $l_{j}=\ln({\alpha_1^{(j)}})$, $Q=\eta_{1R}+\frac{\epsilon_{11}}{4}$, $L=e^{({R_1-\frac{\epsilon_{11}}{2}})}-2$, $\eta_{1R}=k_{1R}(X-2\delta k_{1I}Z)$, $\eta_{1I}=k_{1I}X+\delta (k_{1R}^2-k_{1I}^2)$. Here $X$ and $Z$ are nonlinearity dependent coordinates as given in Eq. (14). When we analyze the above form, it is clear that ICCSs admit different profiles ranging from double-hump and flat-top structures including an asymmetric single-hump as well (but not in exact $sech$ form) for $P_j\neq 0$ which is possible only for $S_1\neq 0$. %The amplitude of such double-hump and flat-top soliton solutions is obtained as $B_j$.\\ 

Here we should emphasis in both solutions (15) and (16) the amplitude of solitons $B_j$ and $C_j$ in $j$-th mode respectively are strongly influenced by the varying nonlinearity $\gamma(z)$. Additionally, the velocity, position and phase of these solitons are also altered by nonlinearity $\gamma(z)$, which are not at all possible in homogeneous solitons where they depend only on the wave vectors $k_1$. So, by properly choosing the arbitrary nonlinearity parameter, one can engineer the resultant solitons, so that they can be utilized for a wider range of applications. Here, we concentrate on some simple soliton management mechanisms such as amplification, compression, oscillation, and tunneling of solitons with appropriate forms of nonlinearities. % given in Eqs. (\ref{nonlinearity}). 

\subsection{Nature and Impact of Nonlinearities}
\begin{figure}[!pht]
	\begin{center}	\includegraphics[width=.55\linewidth]{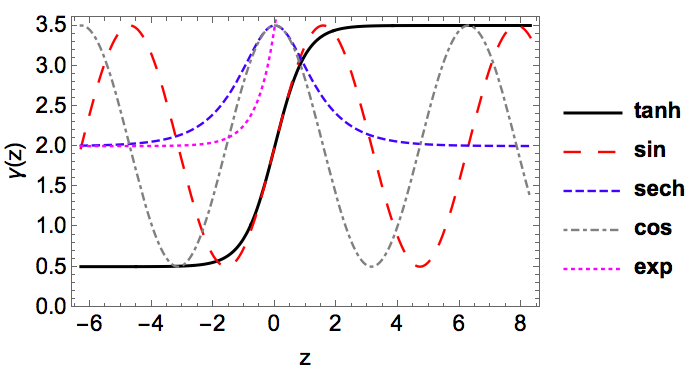}\label{fig1a}~\includegraphics[width=.45\linewidth]{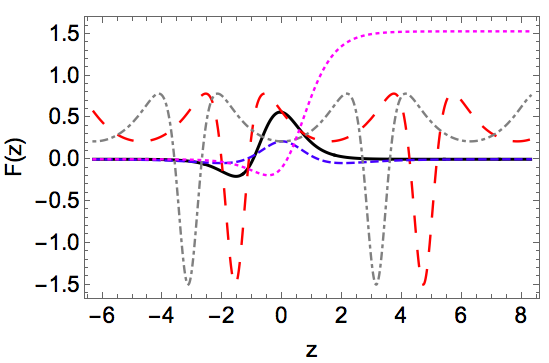}\label{fig1b}
	\end{center}
	\vspace{-0.7cm}
	\caption{Nature of nonlinearity parameter $\gamma(z)$ and $F(z)$ for different choices. (i) tanh: $m=1$ in (17a), (ii) sin: $m=0$ in (17a), (iii) sech: $m=1$ in (17b), (iv) cos: $m=0$ in (17b), and (v) exp: $\gamma_2=0.12$ in (17c) with other values as $\gamma_0=2.0$ and $\gamma_1=1.05$.}
	\label{fig1-non-pot}
\end{figure}

As mentioned below Eq. (12), the nonlinearity $\gamma(z)$ and varying refractive index profile $F(z)$ (ultimately $v(x,z)$)  are mutually dependent. So, throughout our study, we consider a set of interesting nonlinearity functions and explain how they play crucial role in the evolution of solitons in the considered system. Here, we choose the following forms of nonlinearity function as Jacobian elliptic type function (correspond to soliton lattices) and exponential type: %obtain the respective form of $F(z)$ from Eq. ({\ref{potF}). 
\bes \bea
&&\gamma(z)=\gamma_0+\gamma_1 ~\text{sn}(z,m), ~~~~0<m<1,\label{eq-sn}\\
&&\gamma(z)=\gamma_0+\gamma_1 ~\text{cn}(z,m), ~~~~0<m<1,\label{eq-cn}\\
&&\gamma(z)=\gamma_0+\gamma_1 \exp(\gamma_2 z),\label{eq-exp}
\eea\label{nonlinearity}
\ees
where $\gamma_0$, $\gamma_1$, and $\gamma_2$ are arbitrary real constants. From the above nonlinearities, one can obtain explicit form of $F(z)$ from Eq. (12e) which show the nature of varying refractive index profile $v=Fx^2/2$. We portray the variation of these nonlinearities $\gamma(z)$ and potential function $F(z)$ with respect to `$z$' in Fig. \ref{fig1-non-pot}. Especially, for different values of elliptic modulus parameter $m$ in the nonlinearities, they smoothly transfer from a periodic profiles (for $m=0$) to a step-like and localized-hump structures(for $m=1$) in turn alters the refractive index profile too as depicted in the right panel of Fig. \ref{fig1-non-pot}. Such smooth function of nonlinearities are reasonable candidates for experimental implementation \cite{scirep}.  

By adopting the above types of nonlinearities (17), we investigate the evolution of inhomogeneous solitons given by Eqs. (15) and (16) in the following part.

\subsubsection{Creeping or snake-like soliton: Periodic nonlinearity}  
\begin{figure}
	{ \hfill~~~~\qquad$\gamma=\gamma_0+\gamma_1 \mbox{sn}(z,1)$ \hfill  $\gamma=\gamma_0+\gamma_1 \mbox{cn}(z,1)$\qquad \hfill \hfill}\\
	\centering~\includegraphics[width=0.43\linewidth]{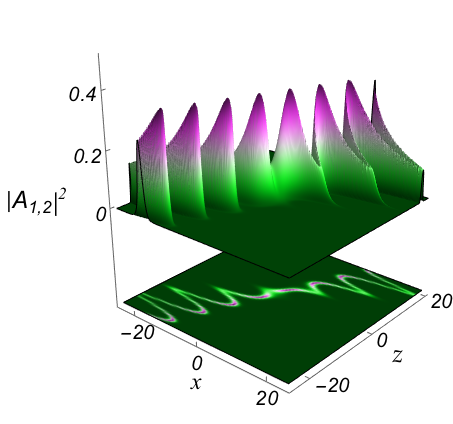}\includegraphics[width=0.43\linewidth]{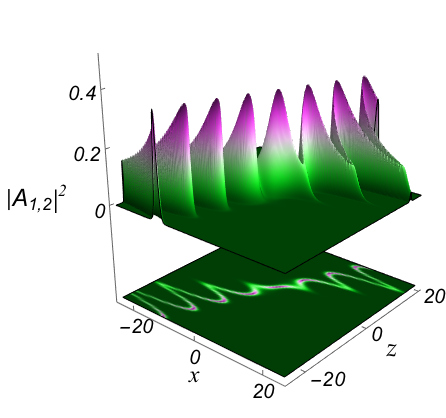}
	\caption{Propagation of single-hump IICSs with periodically oscillating intensity and position/velocity with `sine' and `cosine' type nonlinearities resulting for $m=0$. Both are symmetric to each other except a small shift along '$z$'. Here the parameters are chosen as $\delta=1$, $\epsilon_1=0.5$, and $\epsilon_2=0.12$, $\gamma_0=2.0$, $\gamma_1=1.05$, $k_1=1+0.5i$, $\alpha_1^{(1)}=1.5$, and $\alpha_1^{(2)}=1.5i$.} \label{fig1-sol1-single} \end{figure}

\begin{figure}
	\centering\includegraphics[width=0.43\linewidth]{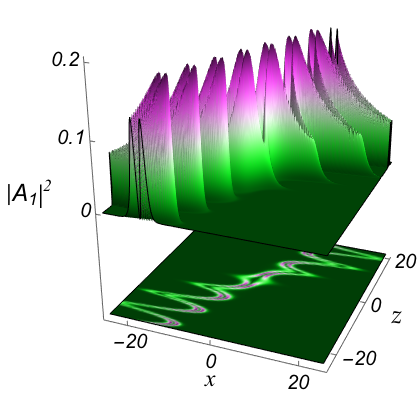}\includegraphics[width=0.43\linewidth]{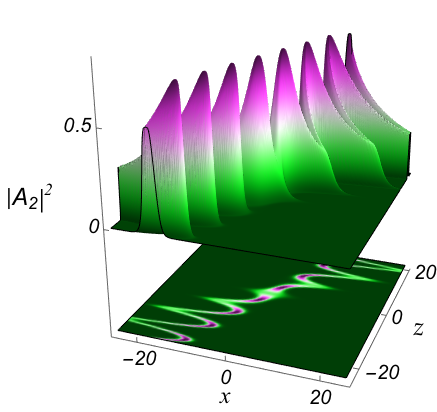}
	\caption{Propagation of double-hump and flat-top ICCSs with periodically oscillating intensity and position/velocity with varying nonlinearity $\gamma=\gamma_0+\gamma_1 \mbox{sn}(z,0)$ for $\alpha_1^{(2)}=2.1i$, while the other parameters as same as in Fig. \ref{fig1-sol1-single}.} 
	\label{fig1-sol1-flat}
\end{figure}

~~\\
Among the considered nonlinearity functions (17), first two forms when $m=0$ leads to the trigonometric type periodic functions namely $\gamma(z)=\gamma_0+\gamma_1 \text{sin}(z)$ and $\gamma(z)=\gamma_0+\gamma_1 \text{cos}(z)$. Mathematically, their implication with respect to $z$ is well known, which is nothing but a periodically oscillating dynamics with a phase-shift between them. In the present system, they affect the nature of soliton propagation such as the amplitude, velocity, and position which become functions of $\gamma(z)$ as evidenced from the explicit solutions (15) and (16). We have shown such traveling soliton in Figs. \ref{fig1-sol1-single} and  \ref{fig1-sol1-flat}, where one can witness the periodically modulated/oscillating amplitude as well as direction/velocity. To be explicit the variation in the amplitude of IICSs can be represented as $B_j= \frac{1}{2}\alpha^{(j)}_1\epsilon_1 \sqrt{\gamma (z)} e^{-R_1/2}$, while that of ICCSs is defined by $C_j=2\epsilon_1\sqrt{\gamma}~\exp\left({({l_j+\delta_{11}^{(j)}-\epsilon_{11}})/{2}}\right)$, with other parameters as shown below Eqs. (15-16), which clearly depicts the significance of nonlinearity parameter $\gamma$ in amplitude. Further, the central position as well as the velocity of IICSs and ICCSs are described by $k_{1R}\epsilon_1\left[\gamma x-2\delta\epsilon_1(\epsilon_1 \epsilon_2+ k_{1I})\int \gamma^2 dz\right]+R_1/2$ and $k_{1R}\epsilon_1\left[\gamma x-2\delta\epsilon_1(\epsilon_1 \epsilon_2+ k_{1I})\int \gamma^2 dz\right]+\frac{\epsilon_{11}}{4}$, respectively. Here the periodic nonlinearity makes the propagation of these localized IICSs/ICCSs resembles the snake-like pattern and it can also be referred as “creeping soliton”. Figure \ref{fig1-sol1-single} depicts the modulation of  IICSs admitting single-hump profile with `sine' and `cosine' type nonlinearities. They have equal-intensity/energy in both components and the main difference between these two nonlinearities is only the well-known phase-shift which can be identified from the substantial shift along '$z$'. Another significant feature is that these can be manipulated with the available arbitrary parameters $\gamma_0$ and $\gamma_1$. By increasing the values of $\gamma_1$ we can control the ``creeping" nature which enhances beating (oscillations) effects in the intensity. Further, we have shown the nature of ICCSs (16) having double-hump and flat-top structures in $A_1$ and $A_2$ components, respectively, in Fig. \ref{fig1-sol1-flat} for `sine' nonlinearity. A similar dynamics can be observed for `cosine' nonlinearity as well which are not shown here. 

\subsubsection{Soliton Amplification and Compression: Kink-like nonlinearity}
\begin{figure} 
	\centering~\includegraphics[width=0.433\linewidth]{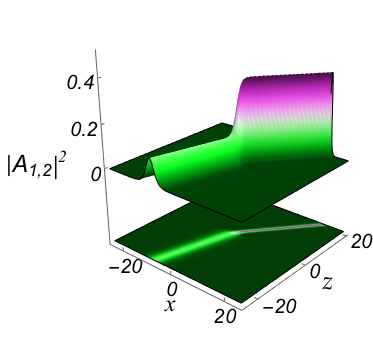}\includegraphics[width=0.433\linewidth]{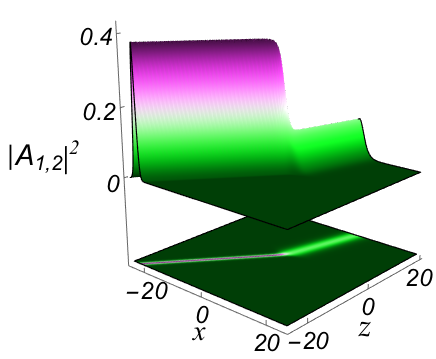} 
	\caption{Controlled amplification and suppression of the intensity associated with pulse compression and widening of single-hump IICSs soliton for $\gamma=\gamma_0+ \gamma_1 \mbox{tanh}(z)$ with $\gamma_1=1.05$ and $\gamma_1=-1.05$, respectively, while the other parameters are $\delta=1$, $\epsilon_1=0.5$, $\epsilon_2=0.12$, $\gamma_0=2.0$, $k_1=1+0.5i$, $\alpha_1^{(1)}=1.5$, and $\alpha_1^{(2)}=1.5i$.} \label{fig-sol1-tanh}
\end{figure}

\begin{figure} 
	\centering\includegraphics[width=0.433\linewidth]{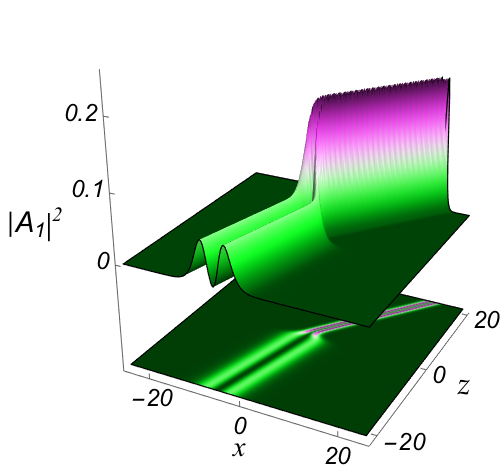}\includegraphics[width=0.433\linewidth]{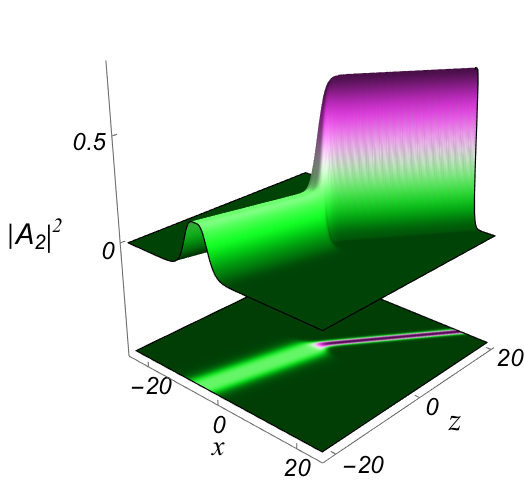}
	\caption{Controlled amplification of the intensity associated with pulse compression of double-hump and flat-top ICCSs with $\gamma=\gamma_0+\gamma_1 \mbox{tanh}(z)$ for the same choice of parameters given in Fig. \ref{fig-sol1-tanh} except for $\alpha_1^{(2)}=2.1i$. }
	\label{fig-sol1-tanh-2}\end{figure}

~~\\
The elliptic function nonlinearity becomes a pure hyperbolic one when $m=1$, especially (17a) turns out to be $\gamma=\gamma_0+\gamma_1\tanh(z)$, and this influences a smooth transition of soliton identities similar to that of a step-function. Soliton dynamics under such `$\tanh$' nonlinearity is shown in Figs.  \ref{fig-sol1-tanh} and \ref{fig-sol1-tanh-2}. Unlike in the previous case of periodic nonlinearities, here the amplitude, velocity, position and width of the solitons are get modulated as a smooth step-like change. To be precise, the intensity of soliton gets amplified with commensurate compression in its width (when $\gamma_1>0$) so that the total energy is conserved during propagation. In contrast when $\gamma_1<0$, the soliton under modulation becomes broader with a suppression/decrease in its intensity. Further, the velocity of the soliton decreases (becomes slower) in the former while it travels faster (increases) in the latter, which consequently changes the actual position of the soliton at any given time. In one way, this clearly indicates the amplitude-independent velocity of solitons in the present system. For a better understanding, we have demonstrated such compressed-amplification and widened-suppression of IICSs in Fig. \ref{fig-sol1-tanh}, where they admit a standard single-hump profile. In the first case with $\gamma_1>0$, the wider soliton having small amplitude (at $z = -20$) gets compressed and undergo significant enhancement in its amplitude (see at $z = 20$). The reverse scenario occurs in the second case with $\gamma_1<0$. Here the shorter/wider soliton travels faster while the taller/narrow soliton propagates slower. To be precise, the kink-like nonlinearity $\gamma(z)$ influences a significant increase in the soliton intensity combined with compression. The nature of modulation in ICCSs under the kink nonlinearity is depicted in Fig. \ref{fig-sol1-tanh-2} which clearly reveals the compressed-amplification of the solitons having double-hump and flat-top profiles with considerable change/reduction in its velocity that forces them to travel slower after the intensity growth. Such phenomenon can be utilized in soliton pulse-shaping dynamics. 

\subsubsection{Soliton Tunneling and Cross-over effects: Bell nonlinearity}
\begin{figure} 
	\centering~\includegraphics[width=0.433\linewidth]{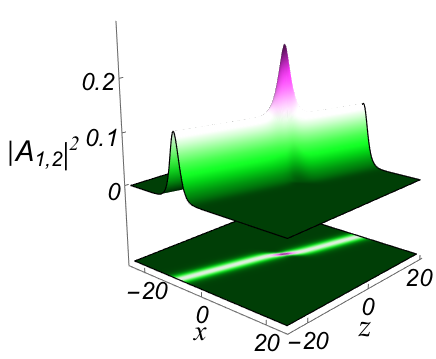}\includegraphics[width=0.433\linewidth]{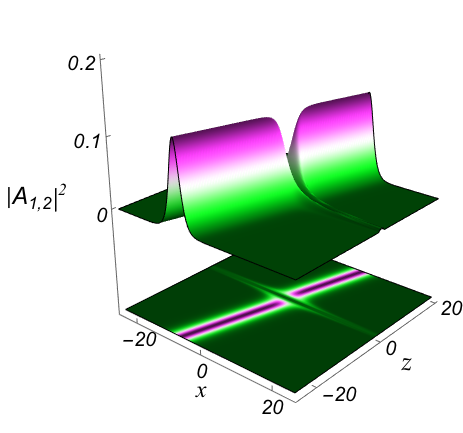}
	\caption{Soliton tunneling and cross-over accompanied by phase-shift through the barrier given by the varying nonlinearity $\gamma=\gamma_0+ \gamma_1 \mbox{sech}(z)$ for $\gamma_1=1.05$ and $\gamma_1=-1.05$, respectively, with $\delta=1$, $\epsilon_1=0.5$, $\epsilon_2=0.12$, $\gamma_0=1.0$, $k_1=1+0.5i$, $\alpha_1^{(1)}=1.5$, and $\alpha_1^{(2)}=1.5i$.}\label{fig-sol1-sech}
\end{figure}

\begin{figure} 
	\centering\includegraphics[width=0.4\linewidth]{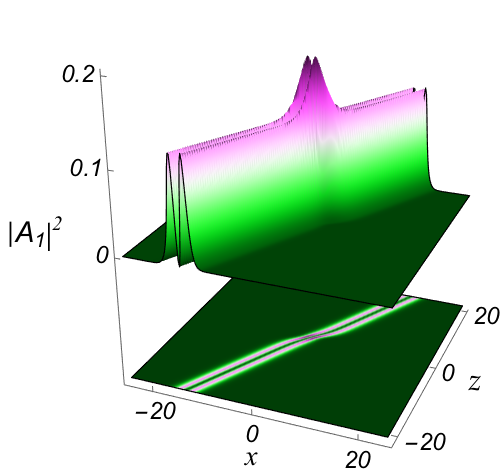}\includegraphics[width=0.4\linewidth]{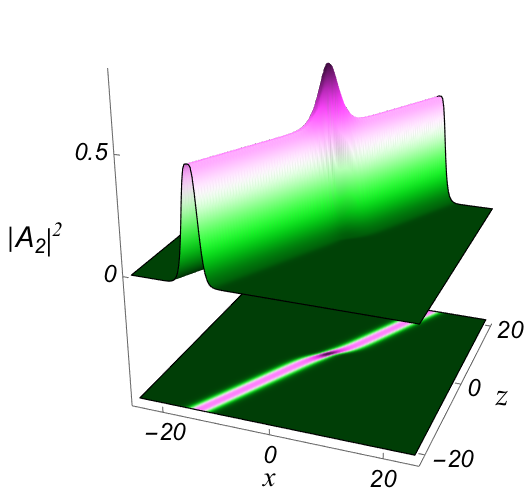}
	\caption{Tunneling of double-hump, and flat-top solitons for $\delta=1$, $\epsilon_1=0.5$, $\epsilon_2=0.12$, $\gamma=1.0+1.05~ \mbox{sech}(z)$, $k_1=1+0.5i$, $\alpha_1^{(1)}=1.5$, and $\alpha_1^{(2)}=2.1i$. Solitons exhibit significant change in their phases after tunneling through the barrier.}
	\label{fig-sol1-sech-2}
\end{figure}

\begin{figure} 
	\centering\includegraphics[width=0.34\linewidth]{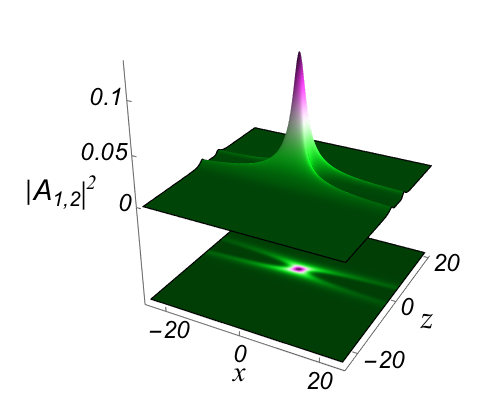}\includegraphics[width=0.33\linewidth]{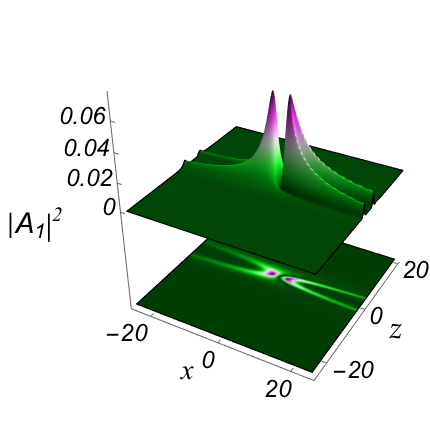}\includegraphics[width=0.33\linewidth]{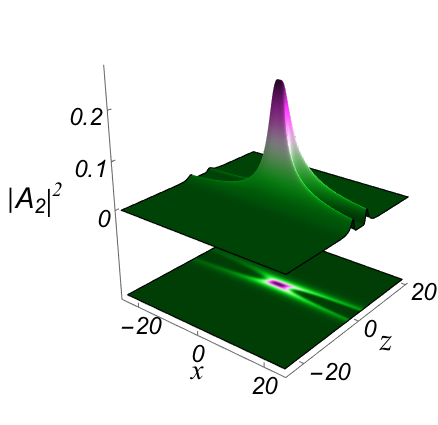}
	\caption{Localized excitation of IICS (single-hump) and ICCSs (double-hump and flat-top solitons) at the barrier for the same type of nonlinearity and arbitrary parameters given in Fig. \ref{fig-sol1-sech-2}, except for $\gamma_0=0.0$.}\label{fig-sol1-sech-3}
\end{figure}

~~\\
Next, we address an interesting concept of soliton tunneling. For this purpose, we consider a nonlinearity of the form (17b) with $m=1$, which leads to a `$\mbox{sech}$' function of `$z$' and in explicit form it can be written as $\gamma=\gamma_0+\gamma_1 \mbox{sech}(z)$. This `$\mbox{sech}$' type nonlinearity creates a localized structure which acts as a barrier. In the present ICCNLS system (10), it gives rise to tunneling of solitons through the barrier as shown in Figs. \ref{fig-sol1-sech} and \ref{fig-sol1-sech-2}. Here the parameter $\gamma_1$ enables the formation of barrier in two categories, one with localized intensity peak (bell type for $\gamma_1>0$) while the other with intensity dip (inverted bell type for $\gamma_1<0$), where the former can be referred as tunneling and the latter as cross-over. An important point to note that is the soliton identities such as amplitude, width and velocity are preserved before and after tunneling. However, there occurs only a small phase-shift due to the tunneling dynamics. Further, when analyze the total energy of the traveling solitons, there occurs compression during the tunneling and widening during cross-over in the barrier regime, see Fig. \ref{fig-sol1-sech}. If we notice the figures clearly, there is no change in the amplitude and width of the soliton until it reaches the barrier and after a short-living compression/widening in the barrier regime, they reemerge with initial characteristics only with a phase-shift. For a complete understanding, we have also demonstrated the ICCSs exhibiting M-shape (double-hump) and flat-top structures for nonlinearity of the form $\gamma=\gamma_0+\gamma_1 \mbox{sech}(z)$ in Fig. \ref{fig-sol1-sech-2}, there itself one can witness the identity preserving tunneling propagation. Such type of tunneling mechanism is looking analogous to the quantum tunneling effect with shape-preservation propagation nature beyond the barrier and they are also referred to soliton spectral tunneling (SST) in the literature \cite{SST}. Such type of tunneling effect appear in a wider context of science including matter wave tunneling in BECs, optical similariton tunneling in photonic crystal fibers, tunneling of self similar optical rogue waves, etc. with an external harmonic potential in both scalar and multicomponent nonlinear systems, see  \cite{7a,tunnel-1,tunnel-2,tunnel-3} and references therein. These prescribe the possibility for observing phenomenon of tunneling experimentally with localized compression. Another interesting observation is the appearance of a rogue-wave-like localized excitations with infinitely long tails when $\gamma_0$ approaches zero, see Fig. \ref{fig-sol1-sech-3}. This phenomenon can be understood in such a way that the solitons vanish along $z$ and emerge/switch as long-lasting tails along $x$. 


\subsubsection{Exponentially growing soliton}
\begin{figure} 
	\centering~\includegraphics[width=0.33\linewidth]{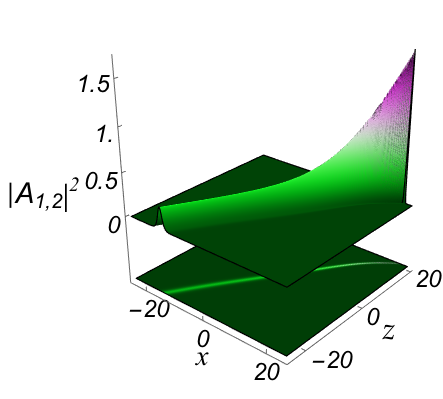}~\includegraphics[width=0.33\linewidth]{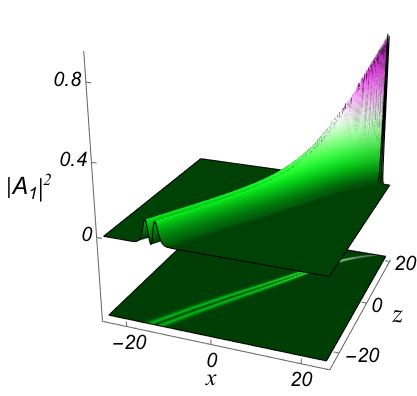}~
	\includegraphics[width=0.33\linewidth]{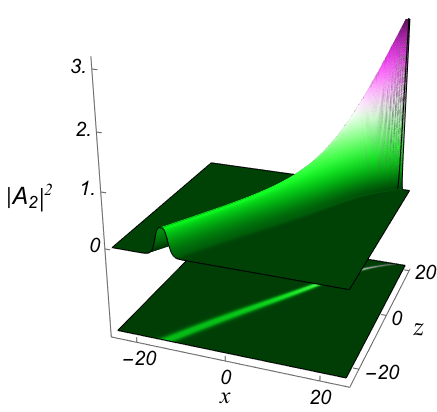}\\
	\includegraphics[width=0.33\linewidth]{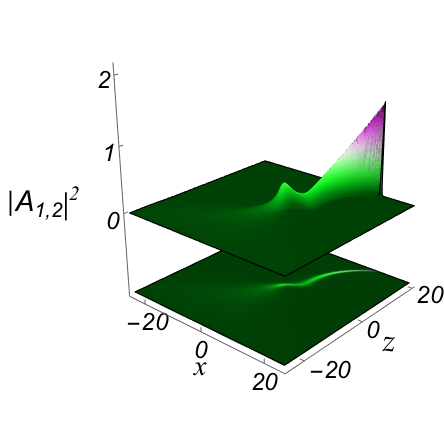}~
	~\quad\includegraphics[width=0.33\linewidth]{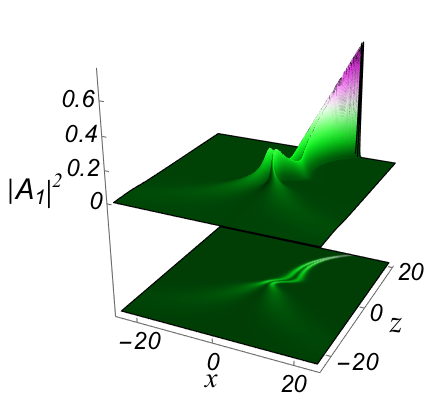}~
	\quad\includegraphics[width=0.33\linewidth]{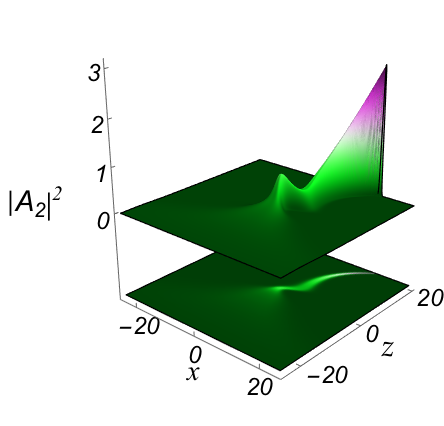}\\
	{\hfill (a) \hfill \hfill(b) \hfill \hfill (c) \hfill }
	\caption{Exponential growth of (a) single-hump IICS and (b-c) ICCSs having double-hump and flat-top profiles associated with compression for exponential (top panel) $\gamma=\gamma_0+\gamma_1 \mbox{exp}(\gamma_2 z)$ and combined (bottom panel) $\gamma=\gamma_0\mbox{sech}z+\gamma_1 \mbox{exp}(\gamma_2 z)$ nonlinearities. The choice of parameters are $k_1=1+0.5i$, $\alpha_1^{(1)}=1.5$, with $\alpha_1^{(2)}=1.5i$ for IICS and $\alpha_1^{(2)}=2.1i$ for ICCS. The other parameters are $\delta=1$, $\epsilon_1=0.5$, $\epsilon_2=0.2$, $\gamma_0=2.0$, $\gamma_1=1.05$, and $\gamma_2=0.12$.} \label{fig-1sol-exp}
\end{figure}

~~\\
In addition to the elliptic function nonlinearities, a classical exponential function also controls the dynamics of soliton propagation in a straightforward rather significant way. Such type of nonlinearities enable the controllable ever-increasing energy of the associated waves. In our case, a simple exponential nonlinearity given by (17c) increases the intensity of the solitons which is well substantiated with compression. For completeness, we have sown such modulation in Fig. \ref{fig-1sol-exp} for $\gamma=\gamma_0+\gamma_1 \mbox{exp}(\gamma_2 z)$. Further, in the presence of superposed nonlinearities with exponential and $\mbox{sech}$ functions, namely $\gamma=\gamma_0\mbox{sech}z+\gamma_1 \mbox{exp}(\gamma_2 z)$, one can understand the soliton growth as well as tunneling dynamics, which we have also depicted in Fig. \ref{fig-1sol-exp}. Additionally, we can observe a strange behaviour of soliton generation from nowhere (for example before $z<-15$) and the intensity picking up when it approaches the barrier and increase exponentially thereafter. Another striking feature is that the width of the solitons are greatly affected by this combined nonlinearity function. A wider (single-hump or double-hump or flat-top) soliton is getting localized when it reaches the barrier and cross-over it, then grow with a very narrow width. %, which can be utilized for 
%%%%%%%%%%%%%%%%%%%%%%%%%%%%%

Here, we wish to remark the integrability and stability of our considered general inhomogeneous model (1). From the obtained similarity transformation (9), equation (1) is turned to become conditionally integrable in the form of equation (10) via Riccati equation (12e). As these inhomogeneous equations are integrable, their solutions also stable. To ensure this, one can follow the stability analysis as presented in Refs. [64,65] which demonstrated how one can identify whether a given solution is stable or unstable on satisfying the conjectured criterion $dP/dv > 0$ and $dP/dv > 0$, respectively,  where $P$ is the normalized momentum and $v$ is the normalized velocity. Proceeding along this direction, we found that the obtained solution fulfils the condition for stability $dP/dv > 0$ and becomes stable. 

%%%%%%%%%%%%%%%%%%%%%%%%%%%%%%%%
\section{Inhomogeneous Two-soliton Solution and their Collisions} \label{sec-2sol}
Being motivated by the effects of inhomogeneous nonlinearities in the one-soliton propagation, here we proceed further to explore their significance in all possible bright soliton collisions of general system (1). For this purpose, first we wish to construct explicit two-soliton solution by following the standard algorithm of Hirota method \cite{Hirota-book} and by using the bilinear form (13) as well as the similarity transformation (9). The bright inhomogeneous two-soliton solution of the general CCNLS system (1) is obtained as 
\bes\bea
\hspace{-1.85cm}A_j(x,z)= \epsilon_1 \sqrt{\gamma} \frac{G^{(j)}}{F} e^{i \zeta(x,z)}\Rightarrow \epsilon_1 \sqrt{\gamma} \frac{G_1^{(j)}+G_3^{(j)}+G_5^{(j)}+G_7^{(j)}}{1+F_2+F_4+F_6+F_8} e^{i \zeta(x,z)}, \quad j=1,2,
\eea where the explicit expression of dependent functions $G^{(1)}=G$, $G^{(2)}=H$, and $F$ takes the following form:
\bea
G_1^{(j)}&=& \alpha _1^{(j)} e^{\eta _1} +\alpha _2^{(j)} e^{\eta _2},\\
G_3^{(j)}&=& e^{2 \eta _1+\eta _1^{*}+\delta_{11}^{(j)}}+e^{2 \eta _1+\eta _2^{*}+\delta_{12}^{(j)}}+e^{2 \eta _2+\eta _1^{*}+\delta_{21}^{(j)}}+e^{2 \eta _2+\eta _2^{*}+\delta_{22}^{(j)}}\nonumber\\
&&+e^{\eta_1+\eta_1^*+\eta_2+\delta_1^{(j)}}+e^{\eta_2+\eta_2^*+\eta_1+\delta_2^{(j)}},\\
G_5^{(j)}&=& e^{2\eta_1+2\eta_1^*+\eta_2+\mu_{11}^{(j)}} +e^{2\eta_1+2\eta_2^*+\eta_2+\mu_{12}^{(j)}}+e^{2\eta_2+2\eta_1^*+\eta_1+\mu_{21}^{(j)}}+e^{2\eta_2+2\eta_2^*+\eta_1+\mu_{22}^{(j)}}\nonumber\\
&&+e^{2\eta_1+\eta_1^*+\eta_2+\eta_2^*+\mu_1^{(j)}}+e^{2\eta_2+\eta_2^*+\eta_1+\eta_1^*+\mu_2^{(j)}},\\
G_7^{(j)}&=&e^{2\eta_1+2\eta_1^*+2\eta_2+\eta_2^*+\phi_1^{(j)}}+e^{2\eta_1+2\eta_2+2\eta_2^*+\eta_1^*+\phi_2^{(j)}},~~~~~~~\\%\eea \bea 
F_2&=&e^{\eta_1+\eta_1^*+R_1}+e^{\eta_1+\eta_2^*+\delta_0}+e^{\eta_2+\eta_1^*+\delta_0^*}+e^{\eta_2+\eta_2^*+R_2},\\
F_4&=&e^{2\eta_1+2\eta_1^*+\epsilon_{11}}+e^{2\eta_1+2\eta_2^*+\epsilon_{12}}+e^{2\eta_2+2\eta_1^*+\epsilon_{21}}
+e^{2\eta_2+2\eta_2^*+\epsilon_{22}}+e^{2\eta_1+\eta_1^*+\eta_2^*+\tau_1}\nonumber\\
&&+e^{2\eta_1^*+\eta_1+\eta_2+\tau_1^*}+e^{2\eta_2+\eta_1^*+\eta_2^*+\tau_2}+e^{2\eta_2^*+\eta_1+\eta_2+\tau_2^*}
+e^{\eta_1+\eta_1^*+\eta_2+\eta_2^*+R_3},\qquad\\
F_6&=&e^{2\eta_1+2\eta_1^*+\eta_2+\eta_2^*+\theta_{11}}+e^{2\eta_1+2\eta_2^*+\eta_2+\eta_1^*+\theta_{12}}+e^{2\eta_2+2\eta_1^*+\eta_1+\eta_2^*+\theta_{21}}\nonumber\\
&&+e^{2\eta_2+2\eta_2^*+\eta_1+\eta_1^*+\theta_{22}},\\
F_8&=&e^{2(\eta_1+\eta_1^*+\eta_2+\eta_2^*)+R_4}.
\eea\label{2sol-solution}\ees
Here $\eta_{j} = k_j(X+i\delta k_j Z)$, $X= ~\epsilon_1 \left({\gamma} x - 2\delta \epsilon_2 \epsilon_1^2\int {\gamma}^2 dz\right)$, and $Z = \epsilon_1^2 \int {\gamma}^2 dz$, while the auxiliary function $S$ utilized in the mathematical process and other quantities are described in the appendix. Note that here $k_j$ represents the wave vectors, while $\alpha_j^{(\ell)}$ denotes the polarization parameters ($j,\ell=1,2$) which are going to play significant role in soliton collisions we discuss in this section.\\

The propagation nature of IICSs and ICCSs under inhomogeneous nonlinearities discussed in the previous section, excites one to explore their collision behaviour driven by the four-wave mixing effect, which is also of considerable attraction. In this connection, the collision of bright solitons in the general CCNLS system (2) can be broadly divided into three categories: (i) IICS $\times$ IICS, (ii) ICCS $\times$ ICCS, and (iii) IICS $\times$ ICCS. Here one should note that the subsystem resulting for the choice (6) supports only the second type of collision between two ICCSs due to the non-availability of IICSs as the choice leads to singular solutions. The remaining two versions of the system exhibit all the above three types of bright soliton collisions. Through a systematic asymptotic analysis and with the aid of already available knowledge on soliton collisions in homogeneous systems, we investigate a categorical analysis on the collision scenario for these three cases, which show elastic and  inelastic/shape-changing type collisions of bright solitons with different profile structures. Mathematically, we identify the form of solitons well before ($z\rightarrow - \infty$) and well after ($z\rightarrow + \infty$) collision, as one could not exactly analyze at the collision point around $z\rightarrow 0$, where the dynamics is quite unpredictable because of its shorter span and nonlinear superposition due to interaction. In a standard way for two-soliton collision, we consider the following asymptotic relations:
\bes\bea 
\hspace{-2.2cm}\mbox{Soliton-1:} \quad \eta_{1R}\approx 0 \Rightarrow \eta_{2R}=2\delta k_{2R}\epsilon_1^2(k_{1I}-k_{2I})\int \gamma^2 dz\approx \pm \infty \quad  \mbox{as} \quad  z\rightarrow \pm \infty,\quad \\
\hspace{-2.2cm}\mbox{Soliton-2:} \quad \eta_{2R}\approx 0 \Rightarrow \eta_{1R}=2\delta k_{1R}\epsilon_1^2(k_{2I}-k_{1I})\int \gamma^2 dz\approx \mp \infty \quad  \mbox{as} \quad  z\rightarrow \pm \infty,\quad
\eea \label{asymp}\ees 
along with the required conditions on $k_{jR}$ and $k_{jI}$ parameters, which we have chosen here as $k_{1R},k_{2R}>0$ and $k_{1I}>k_{2I}$ with opposite signs showing head-on collision (one can also choose same sign to $k_{jI}$ for overtaking collision). 
The mathematical form for asymptotic analyses seems similar to that of homogeneous models reported already \cite{tkjpa,ksjpa,ksjmp} and we do not present their detailed expressions here. Considering the length of the article, we devote this section only for discussion on the above collisions and how they can be controlled/altered by the inhomogeneous nonlinearities. 

\subsection{Elastic Collision of Two IICSs}
\begin{figure} 
	\centering\includegraphics[width=0.314\linewidth]{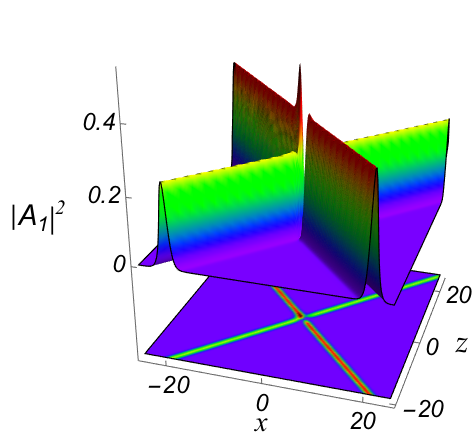}\includegraphics[width=0.33\linewidth]{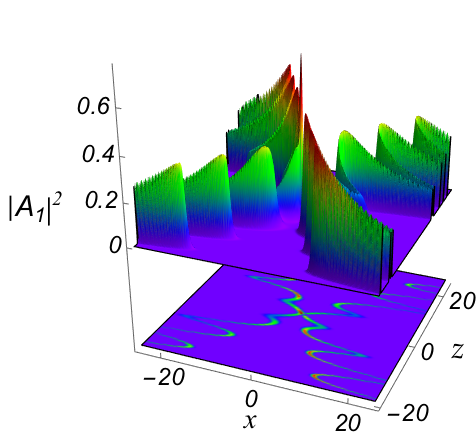}\includegraphics[width=0.33\linewidth]{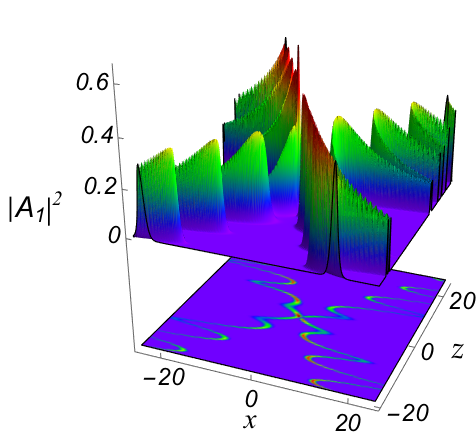}\\
	\includegraphics[width=0.314\linewidth]{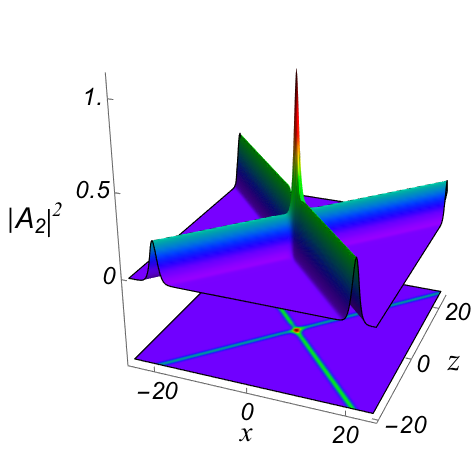}\includegraphics[width=0.33\linewidth]{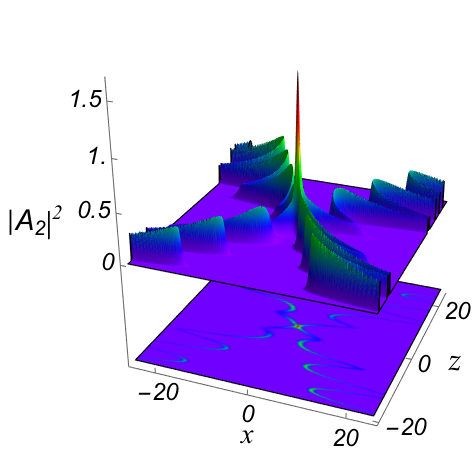}\includegraphics[width=0.33\linewidth]{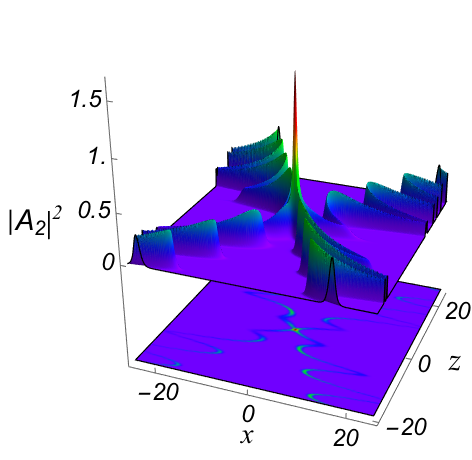}\\
	{\hfill (a) \hfill \hfill(b) \hfill \hfill (c) \hfill }
	\caption{Elastic collisions of two single-hump shaped IICSs under (a) constant nonlinearity and periodic nonlinearities with (b) $\gamma(z)=\gamma_0+\gamma_1 \mbox{sn}(z,0)$ and (c) $\gamma(z)=\gamma_0+\gamma_1 \mbox{cn}(z,0)$. Here the parameters are chosen as $k_1=1+0.5i$, $k_2=1.2-0.5i$, $\alpha^{(1)}_1=0.75$, $\alpha^{(2)}_1=0.75i$, $\alpha^{(1)}_2=0.5$, $\alpha^{(2)}_2=0.5i$, $\delta=1$, $\epsilon_1=0.5$, $\epsilon_2=0.25$, $\gamma_0=2.0$, and $\gamma_1=1.0$. }\label{fig-2sol-case1}
%\end{figure}\begin{figure} 
\centering\includegraphics[width=0.33\linewidth]{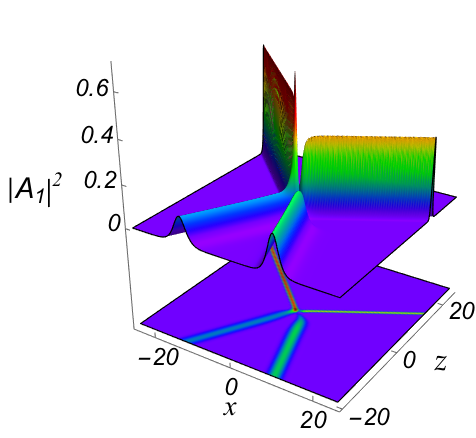}\includegraphics[width=0.33\linewidth]{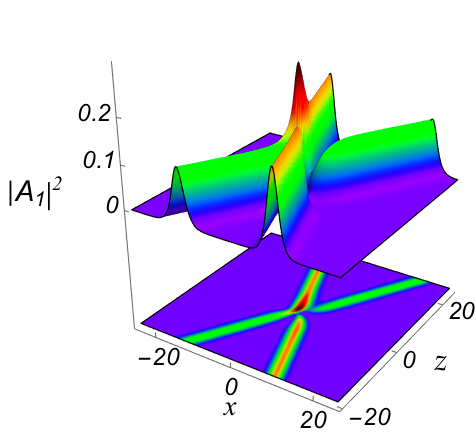}\includegraphics[width=0.33\linewidth]{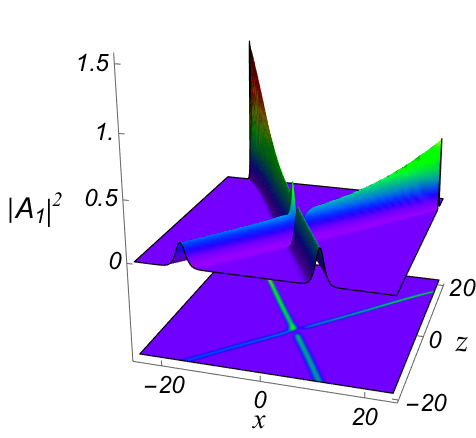}\\ \includegraphics[width=0.33\linewidth]{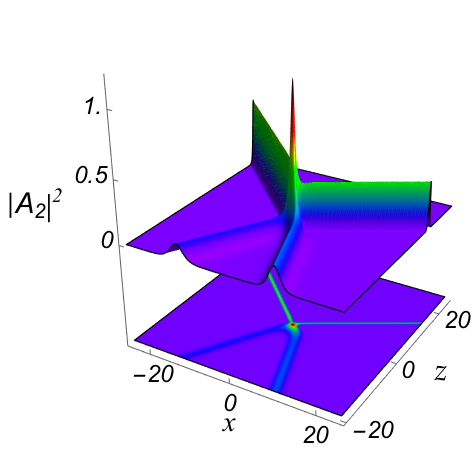}\includegraphics[width=0.33\linewidth]{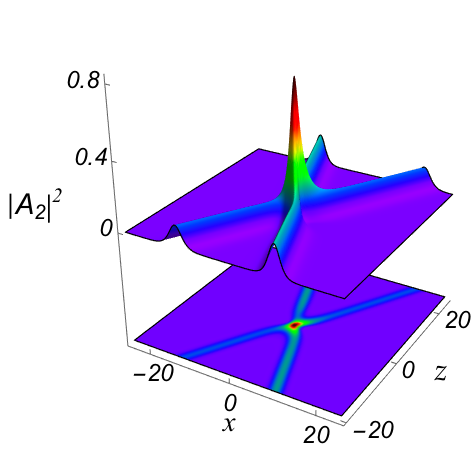}\includegraphics[width=0.33\linewidth]{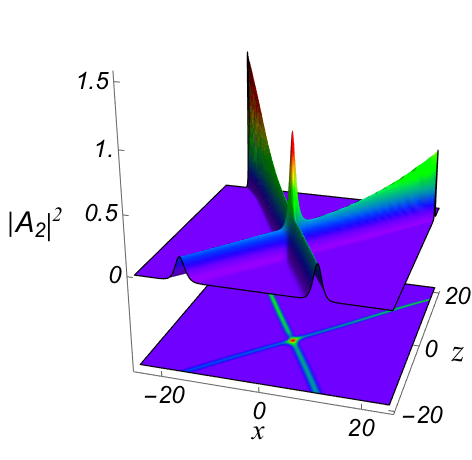}\\
{\hfill (a) \hfill \hfill(b) \hfill \hfill (c) \hfill }
\caption{Elastic collisions of two single-hump shaped IICSs under (a) kink-like nonlinearity $\gamma(z)=\gamma_0+\gamma_1 \mbox{sn}(z,1)$ for $\gamma_0=2.0$, and $\gamma_1=1.0$, (b) bell-type nonlinearity $\gamma(z)=\gamma_0+\gamma_1 \mbox{cn}(z,1)$ for $\gamma_0=1.0$, and $\gamma_1=0.5$, and (c) exponential nonlinearity $\gamma(z)=\gamma_0+\gamma_1 \exp(\gamma_2 z)$ for $\gamma_0=1.5$, $\gamma_1=0.75$ and $\gamma_2=0.1$ with other parameters are chosen as in Fig. \ref{fig-2sol-case1}.}\label{fig-2sol-case1a}
\end{figure}

As mentioned in the previous section, the inhomogeneous incoherently coupled solitons possess a standard single hump profile mathematically represented by a hyperbolic secant function. As the four wave mixing nonlinearity is vanishing for this condition, they behave much similar to the Manakov or two-coupled NLS type solitons. Analysis on their asymptotic dynamics reveals a simple elastic collision between two IICSs by retaining their amplitude, width and velocity after collision, except with a phase-shift. To be explicit, the amplitude of solitons after collision to that of before collision takes the form $B_j^{+}=\frac{(k_1-k_2)(k_2+k_1^*)}{(k_1^*-k_2^*)(k_2^*+k_1)}B_j^-$ for right-moving Soliton-1 and $B_j^{+}=\frac{(k_1^*-k_2^*)(k_2+k_1^*)}{(k_1-k_2)(k_2^*+k_1)}B_j^-$ for left-moving Soliton-2, which ultimately results into $|B_j^{+}|^2=|B_j^-|^2$ representing the unaltered intensities. Here $j=1,2$ denotes the components $A_1,A_2$ and $-/+$ denotes before/after collision. However, these identities are greatly manipulated by the inhomogeneity appearing in the medium which by default affects their collision outcomes as well. For elucidation, we have depicted such a variation imposed by periodic, kink-like, bell-type, and exponentially growing nonlinearities in Figs. \ref{fig-2sol-case1}-\ref{fig-2sol-case1a}. Under constant nonlinearity, both colliding solitons reappears with same amplitude, width and velocity Fig. \ref{fig-2sol-case1}(a). On the other hand, wth periodic type nonlinearities `sn(z,0)' and `cn(z,0)', these identities exhibit periodic variation along temporal direction $z$ both before and after collision. However, the maximum amplitude of both solitons remain same amidst their periodic oscillation which continuously alters the width as velocity as shown in Fig. \ref{fig-2sol-case1}(b-c). 

Further, the effect of kink-like nonlinearity is to enhance the amplitude of solitons accompanied by a compression after collision. This results in an escalated intensity for both the interacting solitons on the same background. Note that the velocity of both solitons are significantly affected (moving too far away from each other). This can be understood from the variation of relative separation distance, which is much larger after collision than that of before collision. This occurs specially for the kink-like nonlinearity only, see Fig. \ref{fig-2sol-case1a}(a). The next bell-type nonlinearity given by $\gamma(z)=\gamma_0+\gamma_1 \mbox{cn}(z,1)$ does not alter the amplitude, velocity or width of the soliton before as well well as after collision except undergoing a tunneling and cross-over effect around the collision point $z\rightarrow 0$ based on the choice $\gamma_1 >0$ and $\gamma_1 <0$, respectively. In Fig. \ref{fig-2sol-case1a}(b), we have shown the tunneling effect in two-soliton collision which have an excess intensity in the tunneling region only. The final form of interest is the familiar one resulting into an exponentially growing intensity of the solitons starting well before the collision and last indefinitely along with commensurate compression of the beam as given in Fig. \ref{fig-2sol-case1a}(c). We have observed a prominent feature of kink-like and exponential nonlinearities, that has the ability to change the nature of soliton collisions. For example, from elastic collision to inelastic collision. In the present case, these nonlinearities changes the elastic collision of two IICSs into an inelastic collision and the intensity of both solitons get increased after interaction with a considerable compression in their width, see Figs. \ref{fig-2sol-case1a}(a) and (c). 

\subsection{Elastic Collision of Two ICCSs}
\begin{figure} 
	\centering\includegraphics[width=0.314\linewidth]{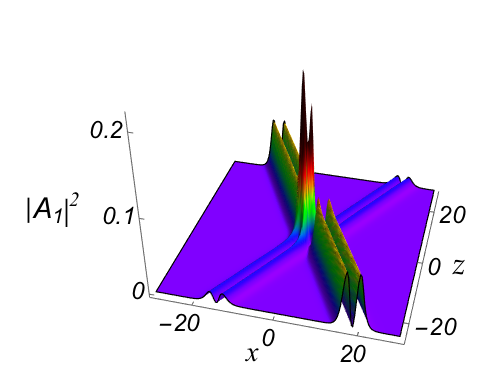}\includegraphics[width=0.33\linewidth]{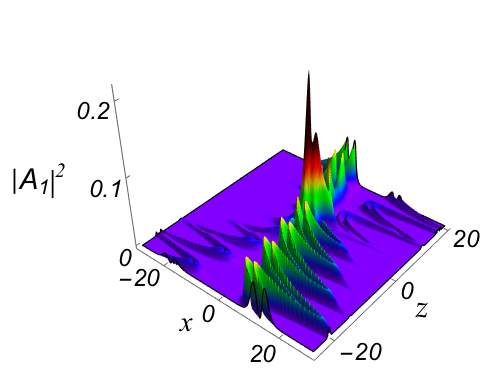}\includegraphics[width=0.33\linewidth]{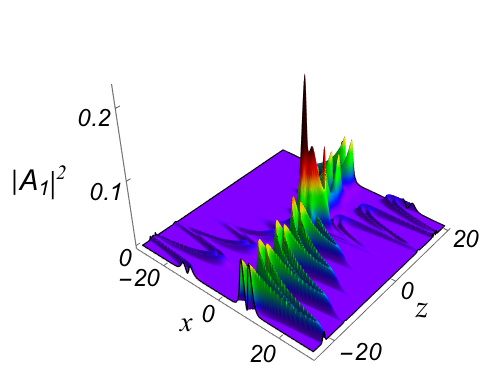}\\
	\includegraphics[width=0.314\linewidth]{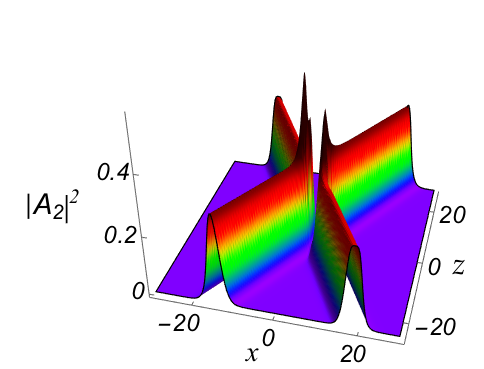}\includegraphics[width=0.33\linewidth]{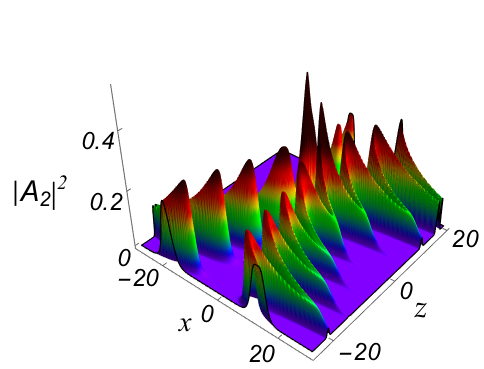}\includegraphics[width=0.33\linewidth]{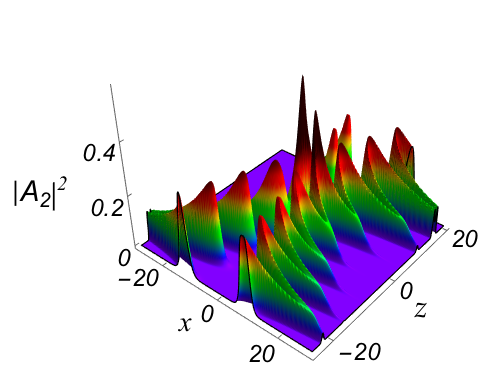}\\
	{\hfill (a) \hfill \hfill(b) \hfill \hfill (c) \hfill }
	\caption{Elastic collisions of two double-hump ICCSs in $A_1$ and collision of single-hump--flat-top ICCSs in $A_2$ components under (a) constant nonlinearity and periodic nonlinearities with (b) $\gamma(z)=\gamma_0+\gamma_1 \mbox{sn}(z,0)$ and (c) $\gamma(z)=\gamma_0+\gamma_1 \mbox{cn}(z,0)$. Here the parameters are chosen as $k_1=1+0.5i$, $k_2=1.25-0.5i$, $\alpha^{(1)}_1=0.75$, $\alpha^{(2)}_1=1.9$, $\alpha^{(1)}_2=1.5$, $\alpha^{(2)}_2=2.1i$, $\delta=1$, $\epsilon_1=0.5$, $\epsilon_2=0.25$, $\gamma_0=2.0$, and $\gamma_1=1.0$. }\label{fig-2sol-case2}
	%\end{figure} \begin{figure} 
	\centering
	\includegraphics[width=0.33\linewidth]{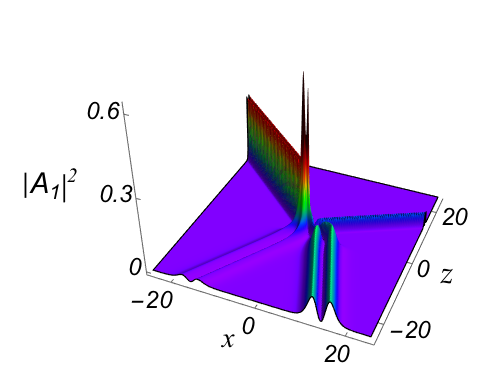}\includegraphics[width=0.33\linewidth]{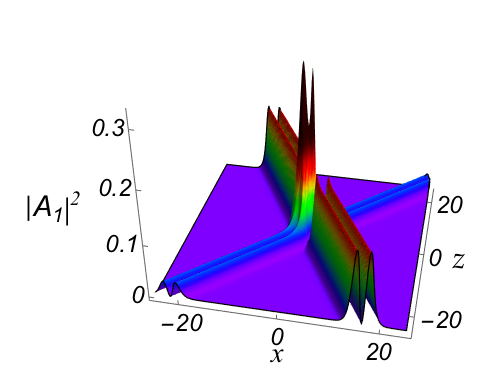}\includegraphics[width=0.33\linewidth]{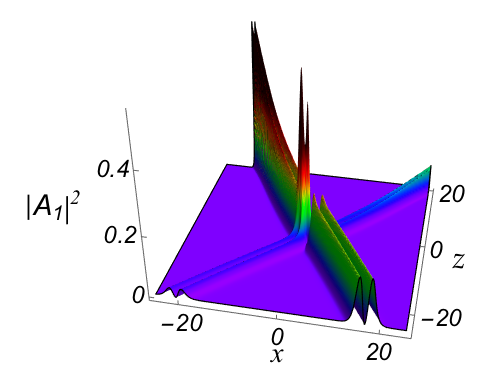}\ \includegraphics[width=0.33\linewidth]{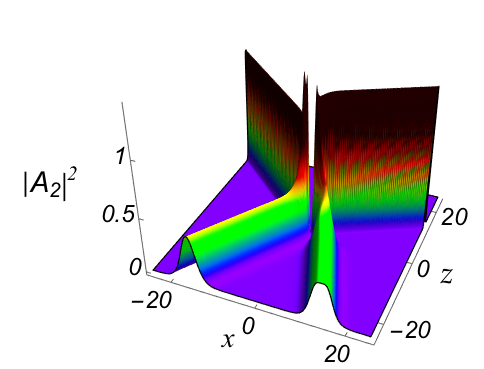}\includegraphics[width=0.33\linewidth]{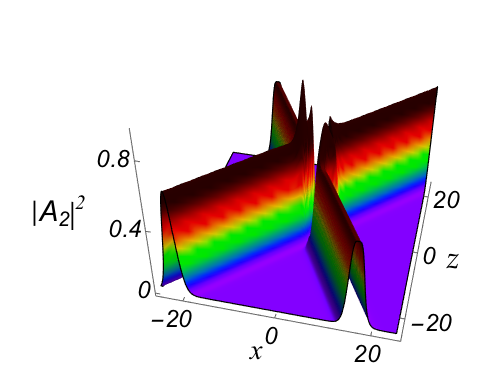}\includegraphics[width=0.33\linewidth]{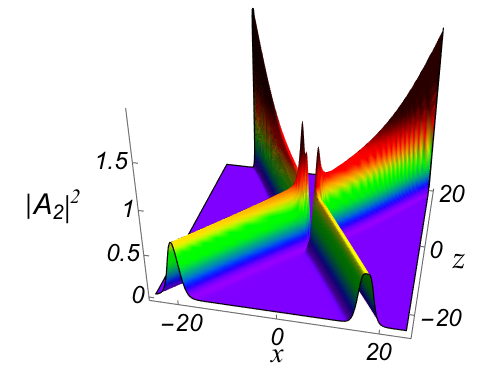}\\
	{\hfill (a) \hfill \hfill(b) \hfill \hfill (c) \hfill }
	\caption{Elastic collisions of two ICCSs with double-hump, flat-top and single-hump structures under (a) kink-like nonlinearity $\gamma(z)=\gamma_0+\gamma_1 \mbox{sn}(z,1)$ for $\gamma_0=2.0$, and $\gamma_1=1.0$, (b) bell-type nonlinearity $\gamma(z)=\gamma_0+\gamma_1 \mbox{cn}(z,1)$ for $\gamma_0=1.5$, and $\gamma_1=-0.5$, and (c) exponential nonlinearity $\gamma(z)=\gamma_0+\gamma_1 \exp(\gamma_2 z)$ for $\gamma_0=1.5$, $\gamma_1=0.5$ and $\gamma_2=0.1$with other parameters are chosen as in Fig. \ref{fig-2sol-case2}.}\label{fig-2sol-case2a}
\end{figure}

When the four-wave mixing nonlinearity comes into picture, we obtain optical soliton profile structures ranging from asymmetric single-hump to symmetric double-hump as well as flat-top. Here, we discuss the collision scenario of such solitons in the presence of inhomogeneous nonlinearities. Our asymptotic analysis shows that the collision between ICCSs is purely elastic as $C_j^{+}=\frac{(k_1-k_2)(k_2+k_1^*)}{(k_1^*-k_2^*)(k_2^*+k_1)}C_j^-$ for right-moving Soliton-1 and $C_j^{+}=\frac{(k_1^*-k_2^*)(k_2+k_1^*)}{(k_1-k_2)(k_2^*+k_1)}C_j^-$ for left-moving soliton which lead to $|C_j^{+}|^2=|C_j^-|^2$. Note that such ICCSs having different profiles reappear unaltered after collision by retaining their shapes along with other identities such as amplitude, width, and velocity. As an example, we have demonstrated such elastic collision of two double-hump ICCSs in the $A_1$ component and collision between a single-hump and flat-top ICCSs in the $A_2$ component in Fig. \ref{fig-2sol-case2}(a). 
As mentioned in the previous section for collision of two IICSs, here these ICCSs undergo modulation by the inhomogeneous nonlinearity without affecting their elastic nature of collision. The sn and cn nonlinearities introduce periodic variation of their amplitudes and velocities with smaller magnitude of temporal oscillations near $z\rightarrow 0$ but with same amplitudes. The kink and bell type nonlinearities do not induce any oscillations in the the amplitude or velocity, instead amplify the intensity after the transition region in the former while the amplification/suppression at the barrier/well occurs in the latter. %Due to increased intensity, it may look like an inelastic collision for the kink-like nonlinearity. However, the width of solitons get compressed correspondingly, thereby preserving the energy of individual solitons. 
This is quite simple in the case of bell-type nonlinearity which by default preserves the intensities. Finally, the exponential nonlinearity increases the intensities throughout their propagation as well as under collision by retaining its elastic nature. Also, here we should note their continuous beam compression which is quite opposite to that of amplification. Similar to the previous case, here also the kink-like and exponential nonlinearities alters the elastic collision of two ICCSs into an inelastic collision. To be precise, both ICCSs collide and exhibit a step/continuous amplification in both components for kink-like/exponential nonlinearity by retaining their profile identities even after collisions, such as both double-hump ICCSs in $A_1$ and a single-hump--double-hump ICCSa in $A_2$. For illustrative purpose and completeness, we have shown such periodically oscillating, step-like compressed amplification, cross-over of the well and amplification with uniform compression of two ICCSs collisions in Figs. \ref{fig-2sol-case2}-\ref{fig-2sol-case2a} for the appropriate choices of arbitrary parameters. 

\subsection{Energy-Switching Collision of  ICCS with IICS}
\begin{figure} 
	\centering\includegraphics[width=0.314\linewidth]{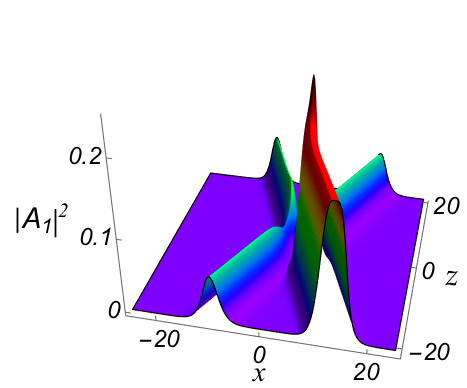}\includegraphics[width=0.33\linewidth]{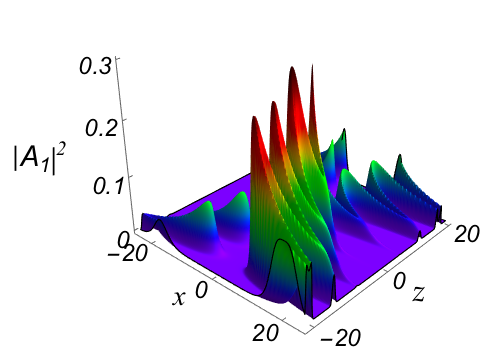}\includegraphics[width=0.33\linewidth]{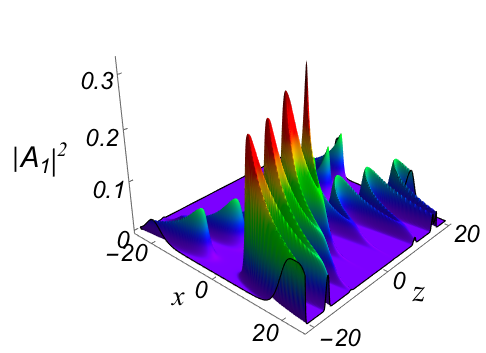}\\
	\includegraphics[width=0.314\linewidth]{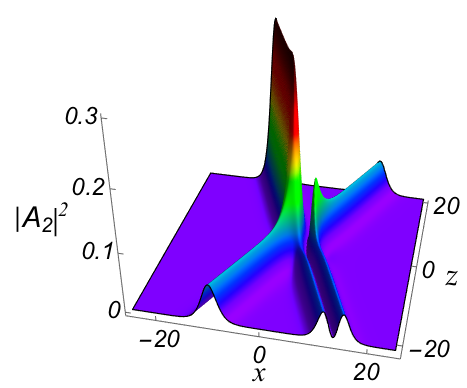}\includegraphics[width=0.33\linewidth]{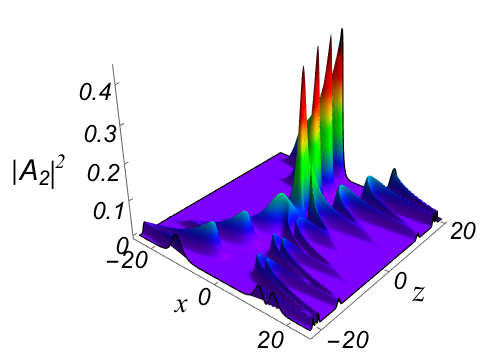}\includegraphics[width=0.33\linewidth]{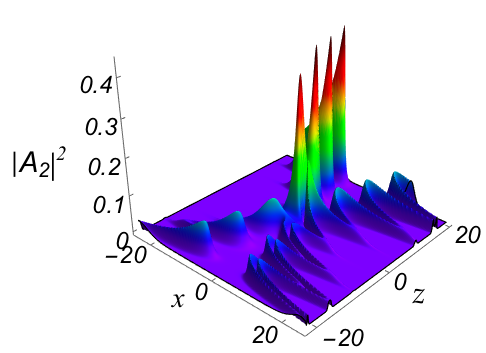}\\
	{\hfill (a) \hfill \hfill(b) \hfill \hfill (c) \hfill }
	\caption{Energy-switching collision of left-moving ICCS due to elastically reappearing right-moving IICS under (a) constant nonlinearity and periodic nonlinearities with (b) $\gamma(z)=\gamma_0+\gamma_1 \mbox{sn}(z,0)$ and (c) $\gamma(z)=\gamma_0+\gamma_1 \mbox{cn}(z,0)$. Here the parameters are chosen as $k_1=1+0.5i$, $k_2=1.2-0.5i$, $\alpha^{(1)}_1=0.75i$, $\alpha^{(2)}_1=0.75$, $\alpha^{(1)}_2=0.75$, $\alpha^{(2)}_2=1.54$, $\delta=1$, $\epsilon_1=0.25$, $\epsilon_2=0.2$, $\gamma_0=2.0$, and $\gamma_1=1.0$. }\label{fig-2sol-case3}
\end{figure}

Compared to the previous two collision scenario, the collision between a IICS and ICCS is turned out to be more interesting. For this purpose, we consider a IICS resulting for the choice $S_1=(\alpha_1^{(1)})^2+\delta (\alpha_1^{(2)})^2=0$ and another soliton ICCS arising for the general choice $S_2=(\alpha_2^{(1)})^2+\delta (\alpha_2^{(2)})^2\neq 0$, where the first is assumed to be right-moving while the second is left-moving one. The detailed asymptotic analysis reveals the amplitude variation due to collision as $B_j^{+}=\frac{(k_1-k_2)(k_2+k_1^*)}{(k_1^*-k_2^*)(k_2^*+k_1)}B_j^-$ for right-moving IICS-1 and $C_j^{+}=\left(\frac{(k_1^*-k_2^*)(k_2+k_1^*) |(\alpha_1^{(j)} \kappa_{22}-\alpha_2^{(j)} \kappa_{12})+\alpha_2^{(j)*}(\alpha_1^{(1)}\alpha_2^{(1)} +\alpha_1^{(2)} \alpha_2^{(2)})/(k_1-k_2)|^2}{(k_1^*-k_2^*)(k_2^*+k_1)\kappa_{22}^2 |\alpha_2^{(j)} |^2 }\right)^{1/2}C_j^-$ for left-moving ICCS-2. This amplitude alteration results into the case  $|B_j^{+}|^2=|B_j^-|^2$ representing the unaltered intensities representing elastic collision of IICS. However, $|C_j^{+}|^2 \neq |C_j^-|^2$, which accounts the important reason for inelastic  collision of ICCS. Thus in both components, IICS reappears without any change in their intensity, while the ICSS undergoes a change in its intensities in opposite sense in both components. For example, an increase in $A_1$ is accompanied by a commensurate decrease in $A_2$ which conserves its intensity as well as the total energy of the system. Thus, it can be inferred that the IICS induces the switching of intensity from one component to another component through ICCS, which leads to the name energy-switching collision. Note that these intensity variations are purely dependent on the wave vectors and polarization parameters, not on the nonlinearity of the system. Such an energy-switching collision between IICS and ICCS with constant nonlinearity is shown in Fig. \ref{fig-2sol-case3}(a), where a single-hump right-moving IICS is undergoing elastic collision in both components and the intensity of right-moving ICCS with flat-top profile decreases to a single-hump in $A_1$, while its intensity is increasing in $A_2$ along with a profile change from double-hump to single-hump. This shows that the extra energy of ICCS after collision in $A_2$ is taken/switched from $A_1$, see Fig. \ref{fig-2sol-case3}(a). 

\begin{figure} 
	\centering\includegraphics[width=0.33\linewidth]{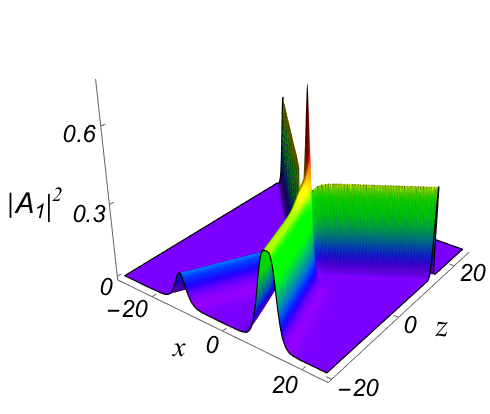}\includegraphics[width=0.33\linewidth]{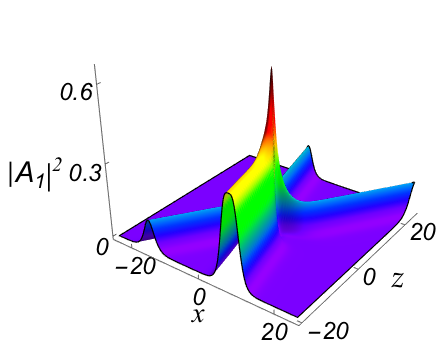}\includegraphics[width=0.33\linewidth]{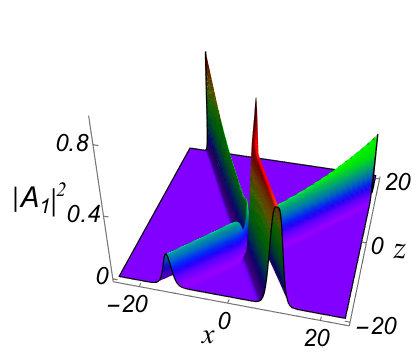}\ \includegraphics[width=0.33\linewidth]{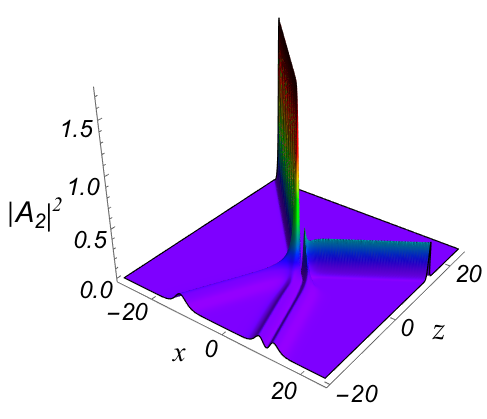}\includegraphics[width=0.33\linewidth]{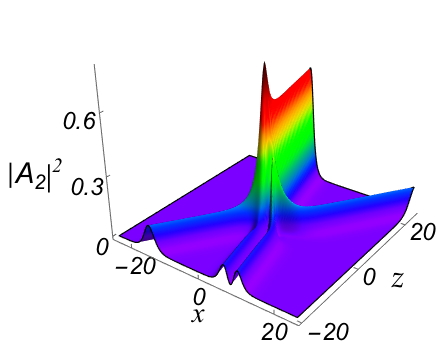}\includegraphics[width=0.33\linewidth]{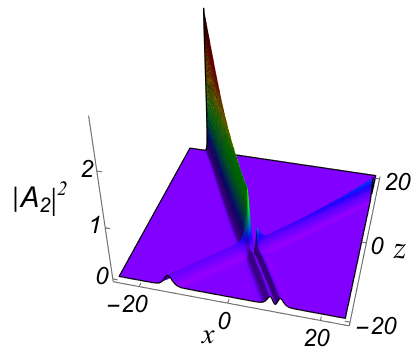}\\
	{\hfill (a) \hfill \hfill(b) \hfill \hfill (c) \hfill }
	\caption{Energy-switching collision of left-moving ICCS due to elastically reappearing right-moving IICS under (a) kink-like nonlinearity $\gamma(z)=\gamma_0+\gamma_1 \mbox{sn}(z,1)$ for $\gamma_0=2.0$, and $\gamma_1=1.0$, (b) bell-type nonlinearity $\gamma(z)=\gamma_0+\gamma_1 \mbox{cn}(z,1)$ for $\gamma_0=1.0$, and $\gamma_1=0.5$, and (c) exponential nonlinearity $\gamma(z)=\gamma_0+\gamma_1 \exp(\gamma_2 z)$ for $\gamma_0=1.5$, $\gamma_1=0.75$ and $\gamma_2=0.075$. Here the other parameters are chosen as in Fig. \ref{fig-2sol-case3}.}\label{fig-2sol-case3a}
\end{figure}

Next, we consider the situation where the nonlinearity is temporally modulated and look for the change in the collision scenario. As discussed earlier, here also, these varying nonlinearities modulate the amplitudes, velocity and width of the participating solitons substantially. The energy-switching nature of ICCS does not change for any choice of considered inhomogeneous nonlinearity, while the IICS is manifesting itself from elastic into an inelastic-switch due to kink-like and exponential nonlinearities in addition to the appropriate modulation in its width through a cascaded compression. For a clear understanding, we have demonstrated energy-switching collision of IICS$\times$ICCS with constant, periodic, kink-like, bell-type, and exponentially varying nonlinearities in Fig. \ref{fig-2sol-case3}(b-c) and Fig. \ref{fig-2sol-case3a}(a-c). As their implications are well discussed in the previous cases, we refrain from giving here again. All these nonlinearities induce change in their identities of the colliding solitons IICS and ICCS, but without affecting the switching nature. Thus, we are getting IICS-ICCS collision with periodical variations, step-like amplitude enhancement along with compression, tunneling through an high amplitude barrier, and continuous amplification with compression. 

\section{Inhomogeneous Soliton Bound States} \label{sec-bound}
\begin{figure}
	\centering\includegraphics[width=0.33\linewidth]{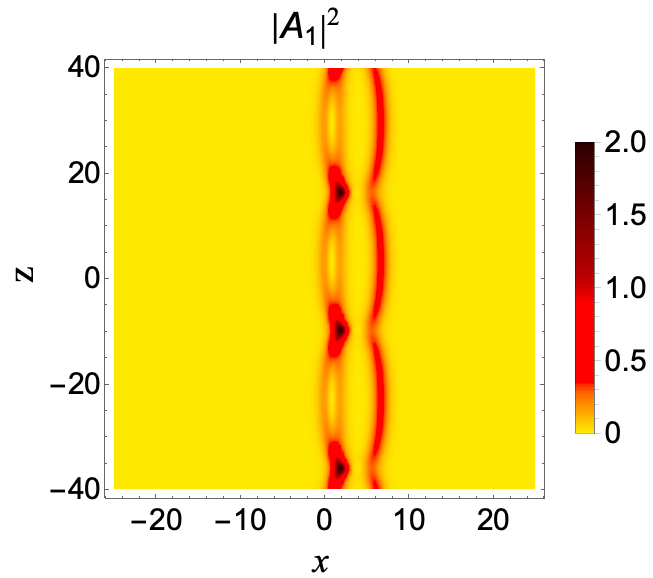}\includegraphics[width=0.33\linewidth]{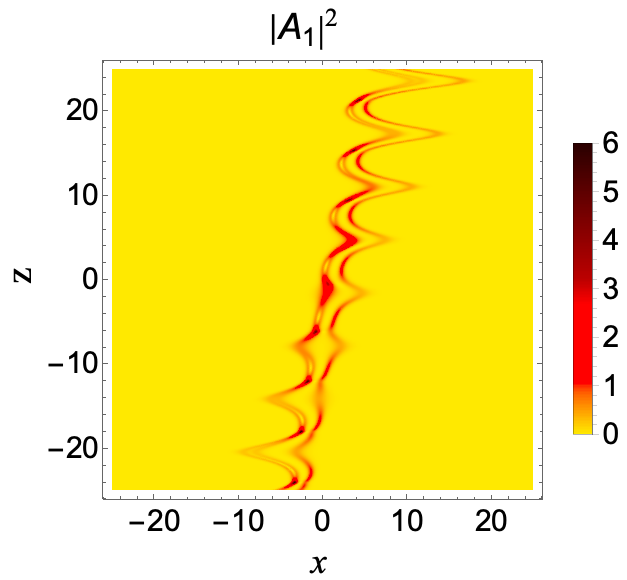}\includegraphics[width=0.33\linewidth]{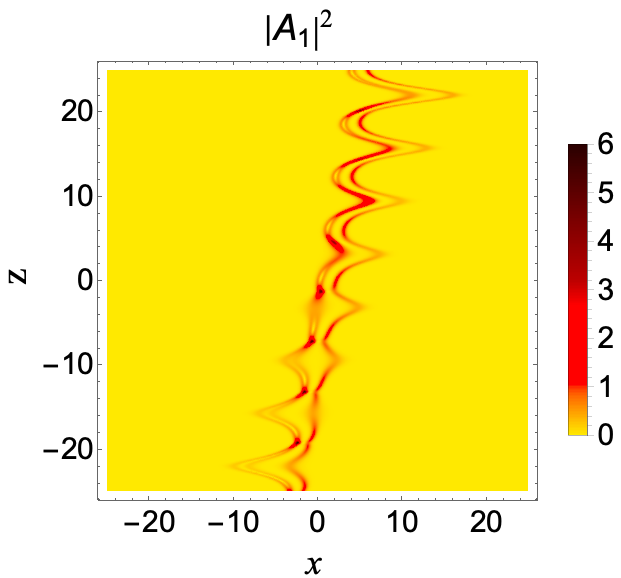}\\
	\includegraphics[width=0.33\linewidth]{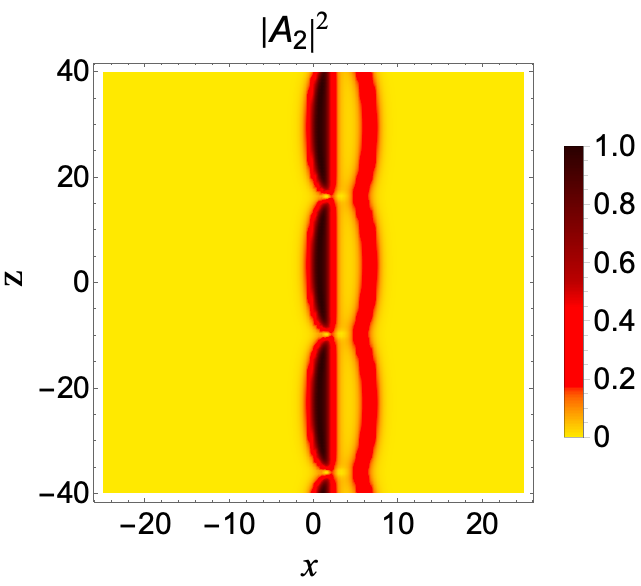}\includegraphics[width=0.33\linewidth]{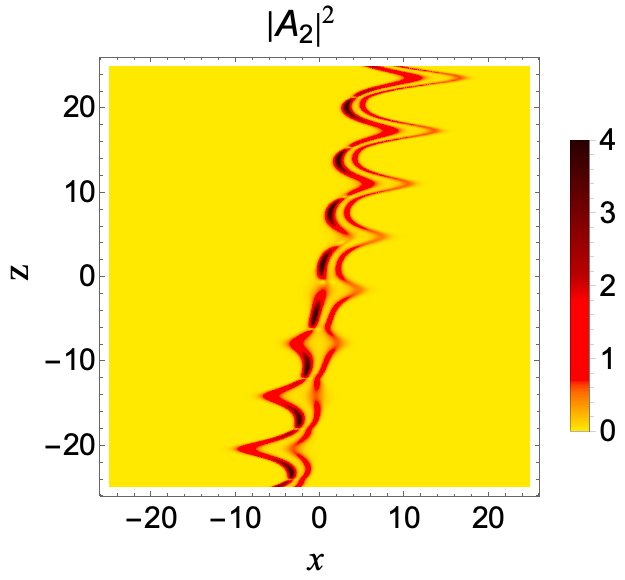}\includegraphics[width=0.33\linewidth]{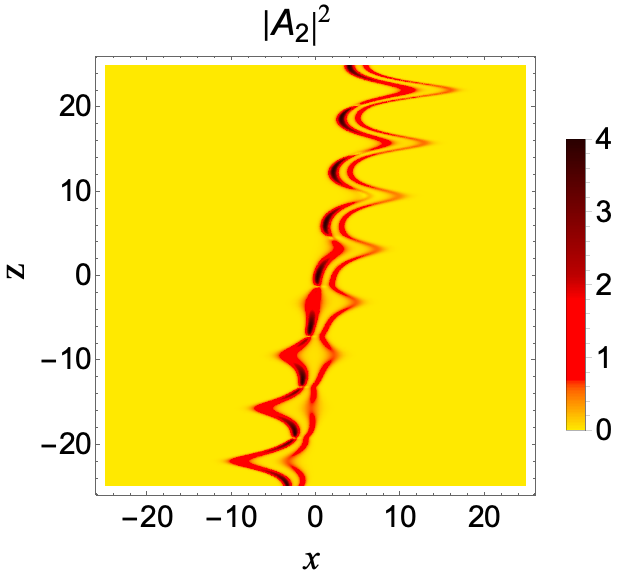}\\
	{\hfill (a) \hfill \hfill(b) \hfill \hfill (c) \hfill \hfill\hfill}\\
	\includegraphics[width=0.33\linewidth]{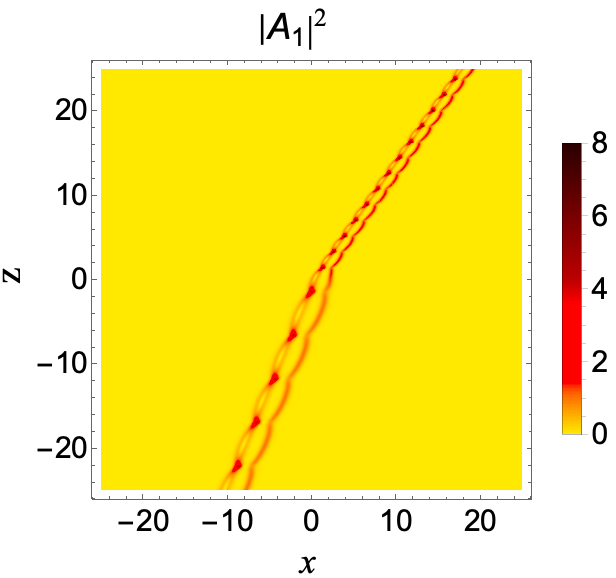}\includegraphics[width=0.33\linewidth]{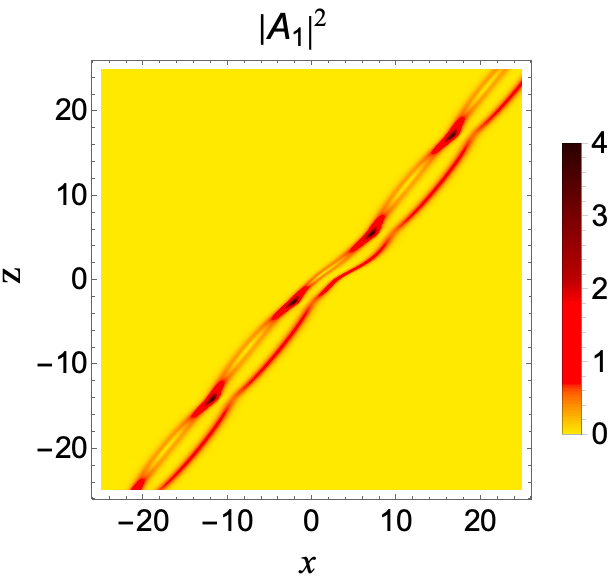}\includegraphics[width=0.33\linewidth]{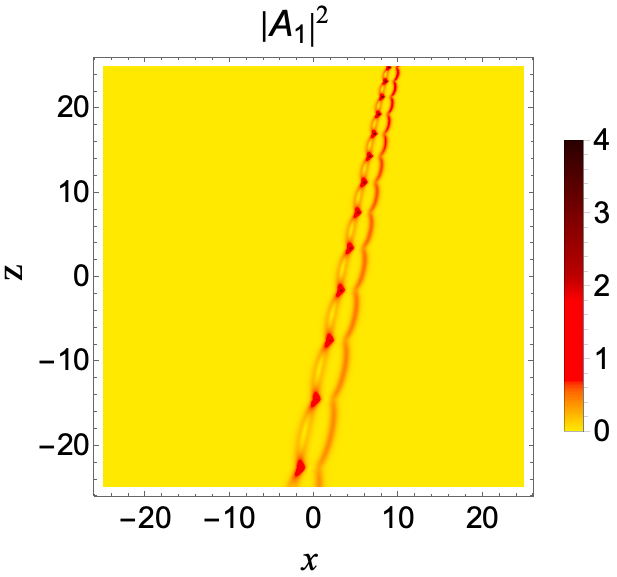}\\
	\includegraphics[width=0.33\linewidth]{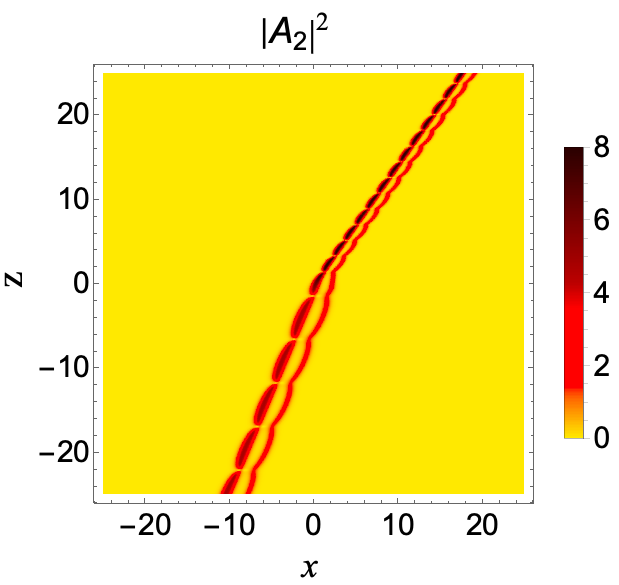}\includegraphics[width=0.33\linewidth]{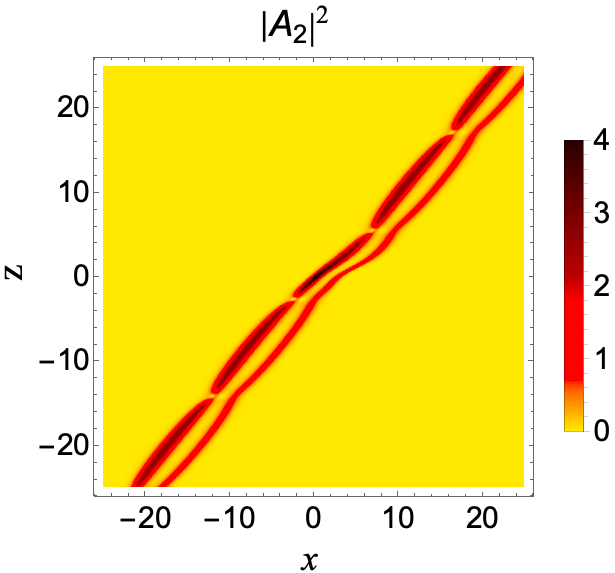}\includegraphics[width=0.33\linewidth]{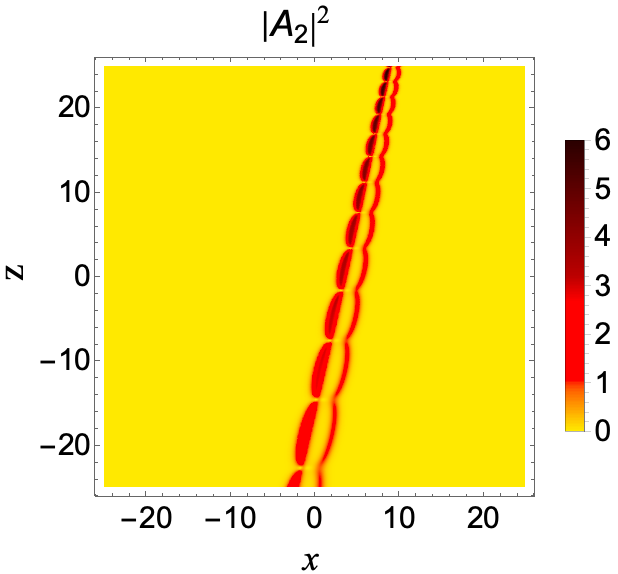}\\
	\centering{\hfill (d) \hfill \hfill(e) \hfill \hfill (f) \hfill }
	\caption{Soliton bound states between two ICCSs having double-hump--single-hump in $A_1$ and flat-top--single-hump in $A_2$ components under (a) constant, (b) periodic-sine, (c) periodic-cosine, (d) kink-like, (e) bell-type, and (f) exponential-growth type nonlinearities. Here the parameters are chosen as (a) $\gamma_0=2.0$, $\gamma_1=\gamma_2=0.0$, $\epsilon_1=0.025$ and $\epsilon_2=0.0$, (b) $\gamma_0=2.0$, $\gamma_1=1.0$, $\gamma_2=-0.12$, $\epsilon_1=0.5$ and $\epsilon_2=0.25$, (c) $\gamma_0=2.0$, $\gamma_1=1.0$, $\gamma_2=-0.12$, $\epsilon_1=0.5$ and $\epsilon_2=0.25$, (d) $\gamma_0=2.0$, $\gamma_1=0.5$, $\epsilon_1=0.75$ and $\epsilon_2=0.25$, (e) $\gamma_0=1.0$, $\gamma_1=0.5$, $\epsilon_1=0.75$ and $\epsilon_2=0.75$, (f) $\gamma_0=1.5$, $\gamma_1=0.75$, $\epsilon_1=0.5$ and $\epsilon_2=0.25$, with other parameters as $k_1=2.3$, $k_2=2.5$, $\alpha_1^{(1)}=0.75i$, $\alpha_1^{(2)}=1.9$, $\alpha_2^{(1)}=1+i$, and $\alpha_2^{(2)}=1-i$. }\label{fig-sbs}
\end{figure}

Further from the soliton collisions, one shall also explore the dynamics of soliton bound states resulting for the choice of equal velocity solitons. In recent years, this attracted much attention in the aspects of soliton molecule formation in optical and atomic systems and referred as velocity resonant solitons. If we consider the case of homogeneous medium with constant nonlinearity and the velocity of two solitons are depend on the wave vectors, in particular their imaginary part $k_{jI}$ for $j$-th soliton. Under such velocity resonance $k_{1I}=k_{2I}$, there occurs a periodic attraction and repulsion of contributing solitons, which are usually called as breathing of solitons. Unlike the standard breathers appearing on constant non-zero background, these breathing solitons can exist on both zero as well as non-zero background. In our system, we can form soliton bound states for the above discussed three cases of collisions. Further, the bound states among different profile solitons are also possible with appropriate choice of polarization parameters determining the contribution of four-wave mixing nonlinearity. Without providing much mathematical forms, we have demonstrated such soliton bound states arising between two ICCSs exhibiting double-hump--single-hump (in $A_1$) and flat-top--single-hump (in $A_2$) structures are shown in Fig. \ref{fig-sbs}. A similar breathing solitons can be observed for the other two cases as well as with non-zero initial velocities  which we have not given here considering the length of the article. Apart from the two-soliton-bound-states, one can investigate the formation of multi-soliton bound structures as well as the interaction between solitons and bound states which is of further interest and we do not discuss the details here. 

\section{Conclusion}\label{sec-conc}
In conclusion, we have investigated the propagation and collision dynamics of inhomogeneous solitons in a system of non-autonomous coherently coupled nonlinear Schr\"odinger (CCNLS) models. By identifying an appropriate similarity transformation  that reduces the considered CCNLS into the canonical integrable CCNLS systems and with the aid of the Hirota’s bilinear
method, we have constructed general soliton solutions. In particular, we classify the solutions into two categories and refer them as inhomogeneous coherently coupled solitons and inhomogeneous incoherently coupled solitons that respectively appear in the presence and absence of four wave mixing effects.  Afterwards, the dynamics of these solitons featuring non-trivial profiles are explored by considering  various temporal modulation of nonlinearities namely, step-like switching nonlinearity, optical lattices and exponential nonlinarity.  Our analysis revealed  that depending upon the nature of nonlinearities the modulated CCNLS system can admit various special localized coherent structures displaying distinct behaviours, like periodically varying solitons, soliton compression  with an unusual hike in its intensity,  tunneling/cross-over effects due to a localized barrier/well,  monotonous amplification with compression of width and a sudden appearance of  a solitonic excitation that grows in amplitude during propagation.  Then, influence of such inhomogeneous nonlinearities on soliton collisions are also investigated briefly  along with soliton bound states. Interestingly, we point out that the collision nature can be altered substantially for certain type of nonlinearity management, namely kink-like nonlinearities. The results presented in this work will be applicable to the studies on engineering optical solitons and their experimental realization towards their controlling mechanism.  Especially, this will have ramifications in the study of soliton propagation in graded refractive index (GRIN) medium. Further, the present optical engineering process can be extended to the setting of  atomic soliton management in pseudo-spinor condensates and in binary condensates with four-wave mixing effects. Now, it is quite natural to look for the influence of such modulations in higher-dimensional nonlinear optical systems featuring vortex solitons, soliton bullets, resonant solitons, lump solitons and dromians. An advantage of our present study it can be straightforwardly extended to multicomponent systems with more than two fields, to name a few multimode propagation in GRIN media with anisotropy, spinor condensates with hyperfine spins $F$ = 1 and 2.

\section*{Acknowledgment} 
The work KS was supported by Department of Science and Technology - Science and Engineering Research Board (DST-SERB), Government of India, sponsored National Post-Doctoral Fellowship (File No. PDF/2016/000547). %KS is grateful to Prof.M. Lakshmanan, Professor of Eminence and DST-SERB Distinguished Fellow, Centre for Nonlinear Dynamics, Bharathidasan University, Tiruchirappalli, India, for his constant support and encouragement. 
The work of TK was supported by the Department of Science and Technology- Science and Engineering Research Board (DST-SERB), Government of India, in the form of a major research project (File No. EMR/2015/001408). 

\section*{Appendix}
The auxiliary function $S$ and other quantities appearing in the two-soliton solution take the following form:
the auxiliary function $s$ is given by
\bea
\hspace{-2.5cm}S=&& S_1 e^{2 \eta_1}+S_2 e^{2 \eta _2}+2S_3 e^{\eta _1+\eta _2}+e^{\eta _1+\eta _1^*+2\eta _2+\lambda _{11}}+e^{\eta _1+\eta _2^*+2 \eta _2+\lambda _{12}}+e^{\eta _2+\eta _1^*+2 \eta_1+\lambda _{21}}\nonumber\\
\hspace{-2.5cm}&&+e^{\eta _2+\eta _2^*+2 \eta _1+\lambda _{22}} +e^{2 \eta _1+2 \eta _1^*+2 \eta _2+\lambda _1}+e^{2 \eta _1+2 \eta _2+2 \eta _2^*+\lambda _2}+e^{2\eta _1+\eta_1^*+2 \eta _2+\eta _2^*+\lambda _3}.\nonumber\\ %\eea\bea
\hspace{-2.0cm}e^{R_u}&=&\frac{\kappa _{uu}}{(k_u+k_u^*)},~~ e^{\delta _0}=\frac{ \kappa _{12}}{(k_1+k_2^*)},~~e^{\delta _{uv}^{(j)}}=\frac{\delta\delta_2 \alpha_v^{(j)*}S_u}{2 (k_u+k_v^*)^2}, ~~e^{\epsilon _{uv}}=\frac{S_u S_v^*}{4 (k_u+k_v^*)^4},\nonumber\\
\hspace{-2.0cm}e^{\delta _u^{(j)}}&=&\frac{\delta\delta_2 \alpha _u^{(j)*} S_3+(k_1-k_2) (\alpha _1^{(j)} \kappa _{2u}-\alpha _2^{(j)} \kappa_{1u})}{(k_1+k_u^*) (k_2+k_u^*)},~~ e^{\tau _u}=\frac{S_u S_3^*}{2 (k_u+k_1^*)^2 (k_u+k_2^*)^2},\nonumber\\ % \eea\bea 
\hspace{-2.0cm}e^{\lambda _{uv}}&=&\frac{(k_1-k_2)^2 \kappa_{uv}~ S_{3-u}}{(k_u+k_v^*)(k_{3-u}+k_v^*)^2},~~
e^{\mu _{uv}^{(j)}}=\frac{  (k_1-k_2)^2 \alpha _{3-u}^{(j)} S_u S_v^*}{4 (k_u+k_v^*)^4 (k_{3-u}+k_v^*)^2},\nonumber\\
\hspace{-2.0cm}e^{\theta _{uv}}&=&\frac{ |k_1-k_2|^4 S_u S_v^*}{{4\tilde{D}}(k_u+k_v^*)^2}  (\alpha _{3-u}^{(1)} \alpha _{3-v}^{(1)*}+\delta \alpha _{3-u}^{(2)} \alpha _{3-v}^{(2)*}),\nonumber\\
\hspace{-2.0cm}e^{\lambda _u}&=&\frac{ (k_1-k_2)^4 S_1 S_2 S_u^*}{4 (k_1+k_u^*)^4 (k_2+k_u^*)^4},~~
e^{\lambda _3}=\frac{ (k_1-k_2)^4 S_1 S_2 S_3^*}{2\tilde{D}},~ e^{R_4}=\frac{  |k_1-k_2|^8 |S_1|^2 |S_2|^2}{16\tilde{D}^2}, \nonumber\\
\hspace{-2.0cm}e^{\phi_u^{(j)}}&=&\frac{\delta\delta_2 (k_1-k_2)^4 (k_1^*-k_2^*)^2 {S_1 S_2 \alpha _{3-u}^{(j)*} S_u^*}} {8\tilde{D} {(k_1+k_u^*)^2(k_2+k_u^*)^2}},~~e^{R_3}= \frac{|k_1-k_2|^2(\kappa_{11}\kappa_{22}-\kappa_{12}\kappa_{21})+ |S_3|^2}{(k_1+k_1^*)|k_1+k_2^*|^2(k_2+k_2^*)}, \nonumber\\ %\eea\bea 
\hspace{-2.0cm}e^{\mu_u^{(j)}}&=&\frac{(k_1-k_2)^2 S_u }{2\tilde{D}} \left(\left[(k_{3-u}+k_1^*)^2+(k_2^*-k_1^*)(k_{3-u}+k_2^*)\right] \alpha _{3-u}^{(j)} \alpha _1^{(j)*} \alpha _2^{(j)*}\right.\nonumber\\
\hspace{-2.0cm}&& \left.+(k_{3-u}+k_1^*) (k_1^*-k_2^*) \alpha _1^{(j)*} (\alpha_{3-u}^{(3-u)} \alpha_2^{(3-u)*}) -(k_1^*-k_2^*) (k_{3-u}+k_2^*) \alpha _2^{(j)*}  (\alpha_{3-u}^{(3-u)} \alpha_1^{(3-u)*}) \right.\nonumber\\
\hspace{-2.0cm}&& \left.+ (k_{3-u}+k_1^*) (k_{3-u}+k_2^*) \alpha_{3-u}^{(j)} (\alpha_1^{(3-u)*} \alpha_2^{(3-u)*}) \right),\nonumber
\eea
where
\bea
S_u&=& (\alpha_u^{(1)})^2 +\delta (\alpha_u^{(2)})^2, \quad S_3=\alpha _1^{(1)} \alpha _2^{(1)}+\delta \alpha _1^{(2)} \alpha _2^{(2)},\nonumber\\
\tilde{D}&=&(k_1+k_1^*)^2(k_1^*+k_2)^2 (k_1+k_2^*)^2 (k_2+k_2^*)^2, \nonumber\\
\kappa_{uv}&=&\left({\alpha_u^{(1)} \alpha_v^{(1)*}+\delta_2 \alpha_u^{(2)} \alpha_v^{(2)*}}\right)/{(k_u+k_v^*)}.\nonumber
\eea
Here $u,v,j,l=1,2$, while $\delta$ and $\delta_2$ take suitable value as given in Eqs. (3), (4), and (6) for the respective CCNLS model.

%\newpage
%\section*{References}


\begin{thebibliography}{30}
	
	\bibitem{kiv-book}
	Kivshar Y S and Agrawal G P 2003 {\it Optical Solitons: From Fibers to Photonic Crystals}, Academic Press, San Diego.
	
	\bibitem{ablowitz}
	Ablowitz M J and Segur H 1981 {\it Solitons and the Inverse Scattering Transform} (SIAM, Philadelphia).
	
	\bibitem{wiley}
	Whitham G B 1974 {\it Linear and Nonlinear Waves} (Wiley, New York).
	
	\bibitem{pana}
	Kevrekidis P G, Frantzeskakis D J and Carretero-Gonz{\'a}lez (Eds.) R 2008
	{\it Emergent Nonlinear Phenomena in Bose-Einstein Condensates: Theory and Experiment} (Springer).
	
	\bibitem{boaris}
	Kartashov Y V, Malomed B A and Torner L 2011 {\it Rev. Mod. Phys.} \textbf{83} 247.
	
	\bibitem{expoptics}
	Hukriede J, Runde D and Kip D 2003 {\it J. Phys. D} \textbf{36} R1.
	
	\bibitem{expoptics1}
	Rohrmann P, Hause A and Mitschke F 2013 {\it Phys. Rev. A} \textbf{87} 043834.
	
	\bibitem{optexp1}
	Centurion M, Porter M A, Kevrekidis P G and Psaltis D 2006 {\it Phys. Rev. Lett.} \textbf{97} 033903. %Nonlinearity Management in Optics: Experiment, Theory, and Simulation
	
	\bibitem{becexp2}
	Davis K B, Mewes M O, Andrews M R, Van Druten N J, Durfee D S,  Kurn D M and W. Ketterle 1995 {\it Phys. Rev. Lett.} \textbf{75} 3969.
	\bibitem{becexp2a}
	Bradley C C, Sackett C A, Tollett J J and  Hulet R G 1995 {\it Phys. Rev. Lett.} \textbf{75} 1687.
	\bibitem{becexp2b}
	Bradley C C, Sackett C A and Hulet R G 1997 {\it Phys. Rev. Lett.} \textbf{78} 985.
	
	\bibitem{bloch}
	Bloch I, Dalibard J and Zwerger 2008 {\it Rev. Mod. Phys.} \textbf{80} 885.
	
	\bibitem{Peyrad}
	Dauxois T and Peyrard M 2006 {\it Physics of solitons} (Cambridge University Press, Cambridge).
	
	\bibitem{1a} 
	Kengne E, Lakhssassi A and Li W-M 2019 {\it Nonlinear Dyn.} \textbf{97} 449.
	\bibitem{1a2}
	Wright L G, Renninger W H, Christodoulides D N and Wise F W 2015 {\it Opt.Exp.} \textbf{23} 3492.
	\bibitem{1a3}
	Arabi C M, Kudlinski A, Mussot A and M. Conforti 2018 {\it Phys. Rev. A} \textbf{97} 023803.
	\bibitem{1a4}
	Wang Y Y, Dai D and Dai C Q 2014 {\it Laser Phys.} \textbf{24} 105402.
	\bibitem{1a5}
	Wu L, Zhang J F, Li L, Tian Q and Porsezian K 2008 {\it Opt. Exp.} \textbf{16} 6352.
	
	\bibitem{1b} 
	Turitsyn S K, Bale B G and Fedoruk M P 2012 {\it Phys. Rep.} \textbf{521} 135.
	
	\bibitem{serkin1} 
	Serkin V N, Hasegawa A and Belyaeva T L 2007 {\it Phys. Rev. Lett}. \textbf{98} 074102.
	\bibitem{serkin2} 
	Serkin V N, Hasegawa A and Belyaeva T L 2012 {\it Phys. Rev. A} \textbf{81} 023610.
	
	\bibitem{1c} 
	Ponomarenko S A and Agrawal G P 2006 {\it Phys. Rev. Lett.} \textbf{97} 013901.
	
	\bibitem{maha} 
	Mahalingam A, Uthayakumar A and Anandhi P 2013 {\it J. Opt.} \textbf{42} 182.
	\bibitem{Abdul} 
	Abdullaev F Kh, Umarov B A, Wahiddin M R B and Navotny D V 2000 {\it J. Opt. Soc. Am. B} \textbf{17} 1117.
	
	\bibitem{2a} 
	Li H and Wang D 2007 {\it J. Mod. Opt.} \textbf{54} 807.
	
	\bibitem{3} Xu S, Xue L, Beli\'c M R and He J R 2017 {\it Nonlinear Dyn.} \textbf{87} (2017) 827.
	
	\bibitem{4} Dai C Q and Zhang J F 2013 {\it Nonlinear Dyn.} \textbf{73} 2049.
	
	\bibitem{5a} Dai C Q and Zhu H P 2013 {\it J. Opt. Soc. Am. B} \textbf{30} 3291.
	\bibitem{5b}
	Yang Z Y, Zhao L C, Zhang T, Li Y H and Yue R H 2010 {\it Phys. Rev. A} \textbf{81} 043826.
	
	\bibitem{6a} Neskorniuk V, Lukashchuk A, Ovchinnikov G, K\"uppers F and Chipouline A 2019 {\it Opt. Lett.} \textbf{44} 2657.
	
	\bibitem{6b} Zhong W-P and Beli\'c M 2010 {\it Phys. Rev. E} \textbf{82} 047601.
	
	\bibitem{6c} Li H, Wang T and Huang D 2005 {\it Phys. Lett. A} \textbf{341} 331.
	
	\bibitem{similariton}
	Ponomarenko S A and Agrawal G P 2007 {\it Opt. Lett.} \textbf{32} 1659.
	%Optical similaritons in nonlinear waveguides
	
	\bibitem{prebabu}
	Babu Mareeswaran R, Charalampidis E G, Kanna T, Kevrekidis P G and Frantzeskakis 2014 {\it Phys. Rev. E} \textbf{90} 042912.
	%Vector rogue waves and dark-bright boomeronic solitons in autonomous and nonautonomous settings
	
	\bibitem{7} Mu\~noz Grajales J C 2016 {\it Adv. Math. Phys.} \textbf{19} 5787508.
	
	\bibitem{7a} Musammil N M, Subha P A and Nithyanandan K 2019 {\it Phys. Rev. E} \textbf{100} 012213.
	
	\bibitem{8} Xiao Y, Zhang J and Duan H 2020 {\it Optik} \textbf{212} 164751.
	
	\bibitem{akhm-book}
	Akhmediev N and Ankiewicz A 1997 Solitons: Nonlinear Pulses and Beams (London: Chapman
	and Hall)
	
	\bibitem{kerrmedia}
	Delque M, Fanjoux G and Sylvestre T 2007 {\it Phys. Rev. E} \textbf{75} 016611.
	
	\bibitem{kivshar}
	Dabrowska-W$\ddot{u}$ster B J, Ostrovskaya E A, Alexander T J and Kivshar Y S 2007 {\it Phys. Rev. A} \textbf{75} 023617.
	
	\bibitem{RK97}
	Radhakrishnan R, Lakshmanan M and Hietarinta J 1997 {\it Phys. Rev. E} {\bf 56} 2213.
	
	\bibitem{tkopt}
	Kanna T and Lakshmanan M 2001 {\it Phys. Rev. Lett} \textbf{86} 5043; {\it ibid} 2003 {\it Phys. Rev. E} \textbf{67} 046617.
	
	\bibitem{tkopt2}
	Kanna T, Lakshmanan M, Dinda P T and Akhmediev N 2006 {\it Phys. Rev. E} \textbf{73} 026604.
	
	\bibitem{tkopt3}
	Vijayajayanthi M, Kanna T and Lakshmanan M 2009 {\it Eur. Phys. J. Special Topics} \textbf{173} 57.
	
	\bibitem{akhm}
	Sukhorukov A and Akhmediev N 1999 {\it Phys. Rev. Lett.} \textbf{83} 4736.
	%Coherent and Incoherent Contributions to Multisoliton Complexes
	
	\bibitem{Park}
	Park Q-H and Shin H J 1999 {\it Phys. Rev. E} \textbf {59} 2373.
	
	\bibitem{tkjpa}
	Kanna T, Vijayajayanthi M and Lakshmanan M 2010 {\it J. Phys. A: Math. Theor.} \textbf {43} 434018.
	
	\bibitem{ksjpa}
	Kanna T and Sakkaravarthi K 2011 {\it J. Phys.~A: Math. Theor.} \textbf{44} 285211.
	
	\bibitem{ksjmp}
	Sakkaravarthi K and Kanna T 2013 {\it J. Math. Phys.} \textbf{54} 013701.
	
	\bibitem{tk2011ncnsd}
	Kanna T and Sakkaravarthi K 2011 {\it Novel double hump solitons in coherently coupled nonlinear Schr\"odinger equations}, Sixth National Conference on Nonlinear Systems and Dynamics, Bharathidasan University, Tiruchirappalli, India. 27--30 January.
	
	\bibitem{tkpla16}
	Babu Mareeswaran R and Kanna T 2016 {\it Phys. Lett. A} \textbf{380} 3244.
	
	\bibitem{Wadati-spin} 
	Ho T L 1998 {\it Phys. Rev. Lett.} \textbf{81} 742.
	
	\bibitem{Wadati-spin2}
	Ieda J, Miyakawa T and Wadati M 2004 {\it Phys. Rev. Lett}. \textbf{93} 194102.
	
	\bibitem{Wadati-spin3}
	Ieda J, Miyakawa T and Wadati M 2004 {\it J. Phys. Soc. Jpn}. \textbf{73} 2996.
	
	\bibitem{Theis}
	Theis M, Thalhammer G, Winkler G K, Hellwig M, Ruff G, Grimm R and Denschlag J H 2004 {\it Phys. Rev. Lett.} \textbf{93} 123001.
	%Tuning the Scattering Length with an Optically Induced Feshbach Resonance
	
	\bibitem{tkwcna}
	Kanna T and Sakkaravarthi K 2012 {\it Bilinearization of three component Gross-Pitaevskii equations using a non-standard approach and soliton solutions}, The 6th International Federation of Nonlinear Analysts Conference, University of Athens, Greece, 25 June -- 1 July.
	
	\bibitem{tkpla14}
	Kanna T, Babu Mareeswaran R and Sakkaravarthi K 2014 {\it Phys. Lett. A} \textbf{378} 158.
	
	\bibitem{Hirota-book} %tkopt,exp,akhm,Hirota-book
	Hirota R 2004 {\it The Direct Method in Soliton Theory}, Cambridge University Press, Cambridge.
	
	\bibitem{scirep}
	Li J, Sun K and Chen X 2016 {\it Sci. Rep.} \textbf{6} 38258.
	%Shortcut to adiabatic control of soliton matter waves by tunable interaction
	
	\bibitem{SST}
	Serkin V N, Vysloukh V A and Taylor J R 1993 {\it Electron. Lett.} \textbf{29} 12-13. %Soliton spectral tunnelling effect
	
	\bibitem{tunnel-1} 
	Barak A, Peleg O, Stucchio C, Soffer A and Segev M 2008 {\it Phys. Rev. Lett.} \textbf{100} 153901. %Observation of soliton tunneling phenomena and soliton ejection, 
	
	\bibitem{tunnel-2} Zhong W P and Belic M R 2010 {\it Phys. Rev. E} \textbf{81} 056604. 
	%Soliton tunneling in the nonlinear Schr\"odinger equation with variable coefficients and an external harmonic potential, 
	
	%\bibitem{Nithi-pre19}Musammil N M, Subha P A and Nithyanandan K 2019 {\it Phys. Rev. E} \textbf{100} 012213.
	%Phase dynamics of inhomogeneous Manakov vector solitons, 
	
	\bibitem{tunnel-3} Dai C Q, Zhou G Q and Zhang J F 2012 {\it Phys. Rev. E} \textbf{85} 016603. %Controllable optical rogue waves in the femtosecond regime, 
	
	\bibitem{stab1} 
	Quintero N R, Mertens F G and Bishop A R 2015 {\it Phy. Rev. E} {\bf 91} 012905. 
	%Soliton stability criterion for generalized nonlinear Schr\"odinger equations, Phy. Rev. E  91 (2015) 012905. 
	
	\bibitem{stab2} 
	Kanna T, Babu Mareeswaran R and Mertens F G 2017 {\it J. Phys. Commun.} {\bf 1} 045005.   
	%Non-autonomous bright solitons and their stability in Rabi coupled binary Bose–Einstein condensates, J. Phys. Commun. 1 (2017) 045005.   
	
	%\bibitem{rogue-ccnls}
	%Sun W R, Tian B, Jiang Y and Zhen H-L 2015 {\it Phys. Rev. E} \textbf{91} 023205.
	%Optical rogue waves associated with the negative coherent coupling in an isotropic medium
	
	%\bibitem{exp} 
	%Kibler B, Fatome J, Finot C, Millot G, Dias F, Genty G, Akhmediev N and Dudley J M 2010 {\it Nature Phys.} \textbf{6} 790.
	%The Peregrine soliton in nonlinear fibre optics
	
	%\bibitem{ragav}
	%Pu H, Law C K, Raghavan S, Eberly J H and Bigelow N P 1999 {\it Phys. Rev. A} \textbf{60} 1463.
	%Spin-mixing dynamics of a spinor Bose-Einstein condensate
	
	%[45] C.Q. Dai, Y.Y. Wang, J.F. Zhang, Nonlinear similariton tunneling effect in the birefringent fiber, Opt. Express 18 (2010) 17548–17554.
	
	%[47] M.S. Mani Rajan, J. Hakkim, A. Mahalingam, A. Uthayakumar, Dispersion management and cascade compression of femtosecond nonautonomous soliton in birefringent fiber, Eur. Phys. J. D 67 (150) (2013) 1–8.
	
	%\bibitem{Nithi-chaos17} N. M. Musammil, K. Porsezian, P. A. Subha, and K. Nithyanandan, Chaos 27, 023113 (2017).
	
	%\bibitem{6} Biswas, A., Milovic, D.: Mathematical Theory of Dispersion-Managed Optical Solitons. Springer, Berlin (2010);\\ V.N. Serkin, A. Hasegawa, T.L. Belyaeva, J. Mod. Opt. 57, 456 (2010).
	
	%\bibitem{Ho}
	%Ho T L and Shenoy V B 1996 {\it Phys. Rev. Lett.} \textbf{77} 3276 \\
	%B.D. Esry, C.H. Greene, J.P.J. Burke and J.L. Bohn {\it Phys. Rev. Lett.} \textbf{78} (1997) 3594.
	
	%\bibitem{Wen}
	%Zhang W, Sun B, Chapman M S and You L 2010 {\it Phys. Rev. A} \textbf{81} 033602.
	
	%\bibitem{Theis1}
	%Hamley C D, Bookjans E M, Behin-Aein G, Ahmadi P and Chapman M S 2009 {\it Phys. Rev. A} \textbf{79} 023617.
	
\end{thebibliography}
\end{document}